\documentclass{aa}
\usepackage{graphicx}
\usepackage{txfonts}
\usepackage{multirow}
\usepackage[usenames,dvipsnames]{color}
\usepackage{xspace}
\usepackage[normalem]{ulem}
\usepackage{multicol}

 \def\mso{\,\mathrm{M}_\odot}

 \def\kms{\, {\rm km}\, {\rm s}^{-1}}

 \def\llso{\log\, L/{\rm L}_\odot \,}
 \def\simle{\mathrel{\hbox{\rlap{\hbox{\lower4pt\hbox{$\sim$}}}\hbox{$<$}}}}
 \def\simgr{\mathrel{\hbox{\rlap{\hbox{\lower4pt\hbox{$\sim$}}}\hbox{$>$}}}}
 \def\msoy{\, \mso~{\rm yr}^{-1}}

\begin{document}
\title{Properties of OB star$-$black hole systems derived from detailed binary evolution models}
   %\subtitle{Predictions from detailed binary evolution models}
\author{N. Langer\inst{1,2}
            \and
            C. Sch\"urmann\inst{1,2}
            \and
            K. Stoll\inst{1}
            \and 
            P. Marchant\inst{3,4}
            \and
            D. J. Lennon\inst{5,6}
            \and
            L. Mahy\inst{3}
            \and
            S. E. de Mink\inst{7,8}
            \and
            M. Quast\inst{1}
            \and
            W. Riedel\inst{1}
            \and
            H. Sana\inst{3}
            \and
            P. Schneider\inst{1}
            \and
            A. Schootemeijer\inst{1}
            \and
            Chen Wang\inst{1}
            \and
            L. A. Almeida\inst{9,10}
            \and
            J. M. Bestenlehner\inst{11}
            \and
            J. Bodensteiner\inst{3}
            \and
            N. Castro\inst{12}
            \and
            S. Clark\inst{13}
            \and
            P. A. Crowther\inst{11}
            \and
            P. Dufton\inst{14}
            \and
            C. J. Evans\inst{15}
            \and
            L. Fossati\inst{16}
            \and
            G. Gr\"afener\inst{1}
            \and
            L. Grassitelli\inst{1}
            \and
            N. Grin\inst{1}
            \and
            B. Hastings\inst{1}
            \and
            A. Herrero\inst{6,17}
            \and
            A. de Koter\inst{8,3}
            \and
            A. Menon\inst{8}
            \and
            L. Patrick\inst{6,17}
            \and
            J. Puls\inst{18}
            \and
            M. Renzo\inst{19,8}
            \and
            A. A. C. Sander\inst{20}
            \and
            F. R. N. Schneider\inst{21,22}
            \and
            K. Sen\inst{1,2}
            \and
            T. Shenar\inst{3}
            \and
            S. Sim\'on-D\'ias\inst{6,17}
            \and
            T. M. Tauris\inst{23,24}
            \and
            F. Tramper\inst{25}
            \and
            Jorick S. Vink\inst{20}
            \and
            Xiao-Tian Xu\inst{1}
            }
\institute{Argelander-Institut f\"{u}r Astronomie, Universit\"{a}t Bonn, Auf dem H\"{u}gel 71, 53121 Bonn, Germany
      \and   %below comes no. 2
    Max-Planck-Institut f\"ur Radioastronomie, Auf dem H\"ugel 69, 53121 Bonn, Germany
      \and   %below comes no. 3
    Institute of Astrophysics, KU Leuven, Celestijnenlaan 200D, 3001 Leuven, Belgium
      \and   %below comes no. 4
    Center for Interdisciplinary Exploration and Research in Astrophysics (CIERA) and Department of Physics and Astronomy, 
    Northwestern University, 2145 Sheridan Road, Evanston, IL 60208, USA
      \and   %below comes no. 5
    Instituto de Astrofisica de Canarias, E-38200 La Laguna, Tenerife, Spain
      \and   %below comes no. 6
    Departamento de Astrofisica, Universidad de La Laguna, E-38205 La Laguna, Tenerife, Spain
      \and   %below comes no. 7
    Center for Astrophysics, Harvard-Smithsonian, 60 Garden Street, Cambridge, MA 02138, USA
      \and   %below comes no. 8
    Anton Pannenkoek Institute for Astronomy, University of Amsterdam, 1090 GE Amsterdam, The Netherlands
      \and   %below comes no. 23
    Departamento de F\'isica, Universidade do Estado do Rio Grande do Norte, Mossor\'o, RN, Brazil
      \and   %below comes no. 24
    Departamento de F\'isica, Universidade Federal do Rio Grande do Norte, UFRN, CP 1641, Natal, RN, 59072-970, Brazil
      \and   %below comes no. 9
    Department of Physics and Astronomy, Hicks Building, Hounsfield Road, University of Sheffield, Sheffield S3 7RH, UK 
      \and   %below comes no. 22
    AIP Potsdam, An der Sternwarte 16, 14482 Potsdam, Germany
      \and   %below comes no. 10
    School of physical sciences, The Open University, Walton Hall, Milton Keynes MK7 6AA, UK 
      \and   %below comes no. 11
    Astrophysics Research Centre, School of Mathematics and Physics, Queen’s University Belfast, Belfast BT7 1NN, UK
      \and   %below comes no. 12
    UK Astronomy Technology Centre, Royal Observatory Edinburgh, Blackford Hill, Edinburgh, EH9 3HJ, UK
      \and   %below comes no. 25
    Space Research Institute, Austrian Academy of Sciences, Schmiedlstrasse 6, A-8042 Graz, Austria
      \and   %below comes no. 13
    Universidad de La Laguna, Dpto. Astrofísica, 38206 La Laguna, Tenerife, Spain 
      \and   %below comes no. 14
    LMU Munich, Universit\"atssternwarte, Scheinerstrasse 1, 81679 M\"unchen, Germany
      \and   %below comes no. 15
    Center for Computational Astrophysics, Flatiron Institute, New York, NY 10010, USA
      \and   %below comes no. 16
    Armagh Observatory, College Hill, Armagh, BT61 9DG, Northern Ireland, UK
      \and   %below comes no. 17
    Zentrum f\"{u}r Astronomie der Universit\"{a}t Heidelberg, Astronomisches Rechen-Institut, 
    M\"{o}nchhofstr. 12-14, 69120 Heidelberg, Germany 
      \and   %below comes no. 18
    Heidelberger Institut f\"{u}r Theoretische Studien, Schloss-Wolfsbrunnenweg 35, 69118 Heidelberg, Germany
      \and   %below comes no. 19
    Aarhus Institute of Advanced Studies (AIAS), Aarhus University, Hoegh-Guldbergs Gade 6B, 8000 Aarhus C, Denmark
      \and   %below comes no. 20
    Department of Physics and Astronomy, Aarhus University, Ny Munkegade 120, 8000~Aarhus~C, Denmark
      \and   %below comes no. 21
    IAASARS, National Observatory of Athens, Vas. Pavlou and I. Metaxa, Penteli 15236, Greece
             }
%             
%    \authorrunning{Langer et al.}         
%    \titlerunning{}
   %\date{Received September 15, 1996; accepted March 16, 1997}
%\date{Received; accepted}
\abstract{
% context heading (optional)
The recent gravitational wave measurements have demonstrated the existence of stellar mass black hole binaries.
It is essential for our understanding of massive star evolution to identify the contribution of binary evolution to the formation of double black holes.}
% aims heading (mandatory)   
{A promising way to progress is investigating the progenitors of double black hole systems and comparing predictions with 
local massive star samples such as the population in 30\,Doradus in the Large Magellanic Cloud (LMC). }
% methods heading (mandatory)
{To this purpose, we analyse a large grid of detailed binary evolution models at LMC metallicity with
initial primary masses between 10 and 40$\mso$,
and identify which model systems potentially evolve into a binary consisting of a
black hole and a massive main sequence star. We then derive the observable properties of such systems,
as well as peculiarities of the OB star component.}
% results heading (mandatory)
{We find that $\sim$3\% of the LMC late O and early B\,stars in binaries are expected to possess a black hole companion,
when assuming stars with a final helium core mass above $6.6\mso$ to form black holes.
While the vast majority of them may be X-ray quiet,
our models suggest that these may be identified in spectroscopic binaries, either by large
amplitude radial velocity variations ($\simgr 50\kms$) and simultaneous nitrogen surface enrichment, or through a moderate radial
velocity ($\simgr 10\kms$) and simultaneously rapid rotation of the OB star. The predicted mass ratios are such that main sequence
companions could be excluded in most cases. A comparison to the observed OB+WR binaries in the LMC, 
Be/X-ray binaries, and known massive BH binaries supports our conclusion.}
% conclusions heading (optional), leave it empty if necessary 
{We expect spectroscopic observations to be able to test key assumptions in our models, with important implications
for massive star evolution in general, and for the formation of double-black hole mergers in particular.}
\maketitle
%
%________________________________________________________________
%
\section{Introduction}
\label{sec:intro}
Massive stars play a central role in astrophysics. They dominate the evolution of star-forming galaxies 
by providing chemical enrichment, ionising radiation and mechanical feedback (e.g., Mac Low \& Klessen 2004, Hopkins et al., 2014,
Crowther et al. 2016).
They also produce spectacular and energetic transients, ordinary and superluminous supernovae, and long-duration gamma-ray bursts
(Smartt 2009, Fruchter et al. 2006, Quimby et al. 2011), which signify the birth of neutron stars and black holes
(Heger et al. 2003, Metzger et al. 2017). 

Massive stars are born predominantly as members of binary and multiple systems (Sana et al. 2012, 2014, 
Kobulnicky et al. 2014, Moe \& Di Stefano 2017).
As a consequence, most of them are expected to undergo strong binary interaction, which drastically alters their evolution
(Podsiadlowski et al., 1992, Van Bever \& Vanbeveren 2000, O'Shaughnessy et al. 2008, de Mink et al. 2013).
On the one hand, the induced complexity is a reason that many aspects of massive star evolution are yet not well understood (Langer 2012, Crowther 2020).
On the other hand, the observations of binary systems provide excellent and unique ways to determine the physical
properties of massive stars (Hilditch et al. 2005, Torres et al. 2010, Pavlovski et al. 2018, 
Mahy et al. 2020) and to constrain their evolution (Ritchie et al. 2010, Clark et al. 2014.
Abdul-Masih et al. 2019a).

Gravitational wave astronomy has just opened a new window toward understanding massive star evolution. 
Since the first detection of cosmic gravitational waves on September 14, 2015 (Abbott et al. 2016), reports about the discovery of such events
have become routine (Abbott et al. 2019), with a current rate of $\sim$1 per week. 
Most of these sources correspond to merging stellar mass black holes with
high likelihood (cf., {\tt https://gracedb.ligo.org/latest/}). It is essential to explore which fraction of these
gravitational wave sources reflects the end product of massive close binary evolution, compared to products of dynamical
(Kulkarni et al. 1993, Sigurdsson \& Hernquist 1993, Antonini et al. 2016, 
Samsing \& D'Orazio 2018, Fragione et al. 2019, Di Carlo et al. 2019) and primordial (Nishikawa et al. 2019) formation paths.   
%near galactic nuclei Fragione+2019
%young star clusters Di Carlo+2019
%globular cluster Samsing + D'Orazio18
%Nishikawa+19 primordial

Two different evolutionary scenarios for forming compact double black hole binaries have been proposed. The first one involves
chemically homogeneous evolution (Maeder 1987, Langer 1992, Yoon \& Langer 2005), which may lead to the avoidance of
mass transfer in very massive close binaries (de Mink et al. 2009) and allows compact main sequence binaries to directly evolve
into compact black hole binaries (Mandel \& de Mink 2016). This scenario has been comprehensively explored through detailed
binary evolution models (Marchant et al. 2016), showing that it leads to double black hole mergers only at low metallicity
($Z\simle Z_{\odot}/10$), and is restricted to rather massive black holes ($\simgr 30\mso$; see also de Mink \& Mandel 2016). 

The second proposed path towards the formation of compact double black hole binaries is more complex, and involves 
mass transfer due to Roche-lobe overflow and common envelope evolution (Belczynski et al. 2016, Kruckow et al. 2018). 
At the same time, this path predicts a wide range of parameters for the produced double compact binaries,
i.e., it resembles those suggested for forming merging double neutron stars (e.g., Bisnovatyi-Kogan \& Komberg 1974, 
Flannery \& van den Heuvel 1975, Tauris et al. 2017), double white dwarfs (Iben \& Tutokov 1984, Webbink 1984)
and white dwarf-neutron star binaries (Toonen et al. 2018). Although this type of scenario has not been verified 
through detailed binary evolution models, there is little doubt that the majority of objects in the observed populations of 
close double white dwarfs (Breedt et al. 2017, Napiwotzki et al. 2020) and double neutron stars 
(Tauris et al. 2017, Stovall et al. 2018, Andrews \& Zezas 2019) have been evolving accordingly. 
Consequently, we may expect that also close double black holes from in a similar way.

%______________________________________________

   \begin{figure}
   \centering
   \includegraphics[width = \linewidth]{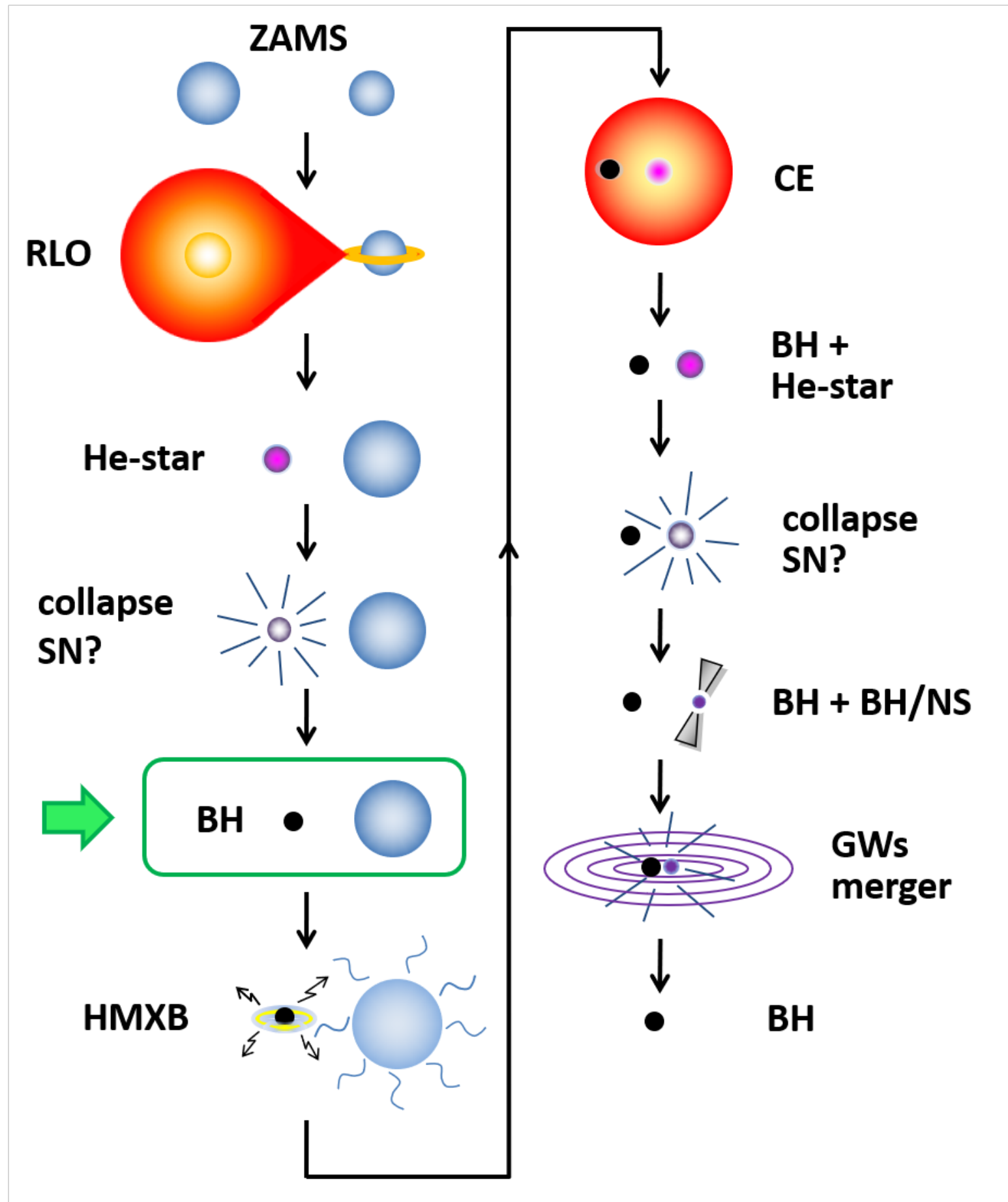}
   \caption{Schematic evolutionary path for the formation of compact BH-NS systems. 
Acronyms used in this figure: ZAMS: zero-age main sequence; RLO: Roche-lobe overflow
(mass transfer); He-star: helium star (could be a Wolf-Rayet star, if sufficiently massive); SN: supernova; BH: black
hole; HMXB: high-mass X-ray binary; CE: common envelope; NS:
neutron star. Light green highlights the OB+BH stage, which is the focus of this paper.
Adapted from Krokow et al. (2018).}
   \label{path}
   \end{figure}
%______________________

Figure\,\ref{path} gives an example for the schematic formation path of double compact binaries (Kruckow et al. 2018).
It involves several stages for which current theoretical predictions are very uncertain, most notably
those of Roche-lobe overflow, common envelope evolution, and black hole formation. 
Evidently, it is desirable to obtain observational tests for as many as possible of the various involved evolutionary stages.
For this, it is important to realise that in many of the steps which are shown in Fig.\,\ref{path}, a large fraction
of the binary systems may either merge or break up, such that the birth rate of double compact systems at the end of the path is several 
orders of magnitude smaller than that of the double main sequence binaries at the beginning of the path. 
Observational tests may therefore be easier for the earlier stages, where we expect many more observational counter parts.

Here, the OB+BH stage, where a BH orbits an O or early B-type star, has a prominent role, concerning theory and observations. From the theoretical perspective,
it is the last long-lived stage which can be reached from the double main sequence stage with detailed stellar 
evolution calculation. Whereas the preceding Roche-lobe overflow phase also bears large uncertainties, it can be
modeled by solving the differential equations of stellar structure and evolution, rather than having to rely on
simple recipes for the structure of the two stars. At the same time, the number of main sequence binaries that are expected to merge 
during the first Roche-lobe overflow phase is typically only about half, such that the number of
OB+BH binaries is expected to be significant.  

In this paper, we describe the properties of OB+BH binaries as obtained from a large grid of detailed binary evolution models.
In Sect.\,\ref{sec:method}, we explain which method was used to obtain our results. Our Sect.\,\ref{sec:results} focuses
on the derived distributions of the properties of the OB+BH binaries, while Sect.\,\ref{sec:uncertain} gives a discussion
of the key uncertainties which enter our calculations. We compare our results with earlier work in Sect.\,\ref{sec:comtheo},
and provide a comparison with observations in Sect.\,\ref{sec:comobs}. In Sect.\,\ref{sec:detect}, we discuss 
observational strategies for finding OB+BH binaries, and in Sect.\,\ref{sec:future}, we consider their future evolution.
We summarise our conclusions in Sect.\,\ref{sec:con}.

\section{Method \label{sec:method}}

Our results are based on a dense grid of detailed massive binary evolution models (Marchant 2016). 
These models were computed with the stellar evolution code MESA (Modules for Experiments in Stellar Astrophysics,
Version No.\,8845) 
with a physics implementation as described by Paxton et al. (2015).
{All necessary files to reproduce our MESA simulations are available at 
https://doi.org/10.5281/zenodo.3698636.}

In particular, differential rotation and magnetic angular momentum transport are included as in  
Heger et al. (2000, 2005), with physics parameters set as in Brott et al. (2011). 
Mass and angular momentum transfer is computed according to Langer et al. (2003) and Petrovic et al. (2005), 
and the description of tidal interaction follows Detmers et al. (2008).
Convection is modeled according to the standard Mixing Length Theory (B\"ohm-Vitense 1958)
with a mixing length parameter of $\alpha_{\rm MLT}=1.5$. 

Semiconvection 
is treated as in Langer (1991), i.e., using $\alpha_{\rm SC}=0.01$. {We note that recent evidence may favour
higher values of this parameter, which
could lead to a nuclear timescale post-main sequence expansion to the RSG stage of massive low-metallicity 
stars in a limited mass range (Schootemeijer et al. 2019, Higgins \& Vink 2020, Klencki et al. 2020).
The consequences of this for massive binary evolution will need to be explored (cf., Wang et al., 2020).
It could lead to the prediction of a significant sub-population of Roche-lobe filling, X-ray bright,
B- and A-type supergiant BH binaries (Quast et al. 2019, Klencki et al. 2020), which, especially at low metallicity, 
appears not to be observed. Clearly, more work is needed to clarify the situation. }

Thermohaline mixing 
is performed as in Cantiello \& Langer (2010), and convective core overshooting is applied
with a step-function extending the cores by 0.335 pressure scale heights (Brott et al. 2011). 
However, overshooting is only applied to layers which are chemically homogeneous. This implies
that mean molecular weight gradients are fully taken into account in the rejuvenation process
of mass gaining main sequence stars (cf., Braun \& Langer 1995). 
The models are computed with the same initial chemical composition as those of Brott et al. (2011),
i.e., taking the non-Solar abundance ratios in the LMC into account. Differently from Brott et al.,
here custom-made OPAL opacities (Iglesias \& Rogers 1996) in line with the adopted initial abundances
have been produced and included in the calculations. 

The masses of the primary stars range from 10 to 39.8$\mso$ in steps of 
$\log\left(M_1 / \mso\right) = 0.050$.
For each primary mass, systems with different initial mass ratios 
$q_i = M_2/M_1$ ranging from 0.25 to 0.975, in intervals of 0.025 
are computed and for each mass ratio, there are models with orbital periods 
from 1.41 to 3160\,d in steps of $\log\left( P_i /{\rm d} \right) =0.025$.
The grid consists of a total of 48240 detailed binary evolution models.
Binaries with initial periods below $\sim 5\,$d (for a primary mass of 10$\mso$) 
and $25\,$d (for a primary mass of 39.8$\mso$) undergo 
mass transfer while both stars fuse hydrogen in their cores (Case\,A systems) while most larger period
binaries undergo mass transfer immediately after the primary leaves the main sequence (Case\,B systems).
For higher primary masses, envelope inflation due to the Eddington limit 
(Sanyal et al. 2015) would 
prevent stable Case\,B mass transfer to occur (cf., Sect.\,\ref{sec:uncertain}).
{Figure\,2 gives an overview of the evolutionary end points obtained for models with an initial primary
mass of $\sim 25.12\mso$, with examples for other primary masses provided in Appendix\,B.}

Our models are computed assuming tidal synchronisation at zero age,
which avoids introducing the initial rotation rate of both stars as additional parameters. 
While this is not physically warranted,
it is justified since moderate rotation is not affecting the evolution of the individual stars
very much (Brott et al., 2011; Choi et al. 2016), and the fastest rotators may be binary evolution
products (de Mink et al., 2013; Wang et al. 2020). Moreover, the initially closer binary models (typically
those of Case A), evolve quickly into tidal locking (de Mink et al. 2009), independent of the initial stellar spins.
Also the spins of the components of all post-interaction binaries, in particular those of the OB+BH binaries
analysed here, are determined through the interaction process, where the mass donor fills its Roche-volume
in synchronized rotation also in Case B systems, and the mass gainer is spun-up by the accretion process.

The evolution of our models is stopped if mass overflow at the outer Lagrangian point L2 occurs {(purple color in Fig.\,2)}
in contact binaries {(black hatching in Fig.\,2)}, 
which are otherwise modeled as in Marchant (2016). We further stop the evolution
if inverse mass transfer occurs from a post main sequence component
{(yellow color in Fig.\,2)},
or if a system exceeds the upper mass-loss rate limit {(green color in Fig.\,2)}. 
Any of these condition is assumed to lead to a merger.
%(cf., Fig.\,\ref{fig:grid25}).
Here, the upper mass loss rate limit is set by the condition that the energy 
required to remove the emitted fraction of the transfered matter
exceeds the radiated energy of both stars. 
%Binary models that violate this condition
%are marked with light-blue colour in Fig.\,\ref{fig:grid25}. 
Models surpassing the weaker condition
that the momentum required to remove the non-accreted mass exceeds their photon momentum are assumed
to survive as binaries. 
%they correspond to those dark-blue pixels in Fig.\ref{fig:grid25} which are marked by hatching. 
The systems are evolved at least until central helium depletion of the mass gainer,
while those with helium core masses smaller than $13\mso$ are followed until
core carbon depletion. 

{In the systems with the largest initial orbital periods, the mass transfer rate grows on near-dynamical
time scales to very large values, with a classical common envelope evolution to follow  (red color in Fig.\,2)}.
In some systems, in particular those with the largest initial periods and the most massive secondary stars, 
a merger as consequence of the common envelope evolution may be avoided. {Here, we assume that also these systems
merge}, such that the numbers and frequencies of OB+BH systems that we obtain below must be considered
as lower limits. The systems which survive a common envelope evolution would likely contribute to the
shortest period OB+BH binaries. As such, they would likely evolve into an OB\,star-BH merger later-on,
and not contribute to the production of double compact binaries. More details about the binary evolution 
grid can be found in Marchant (2016).

\begin{figure}[!htbp]
   \centering
   \includegraphics[width = \linewidth]{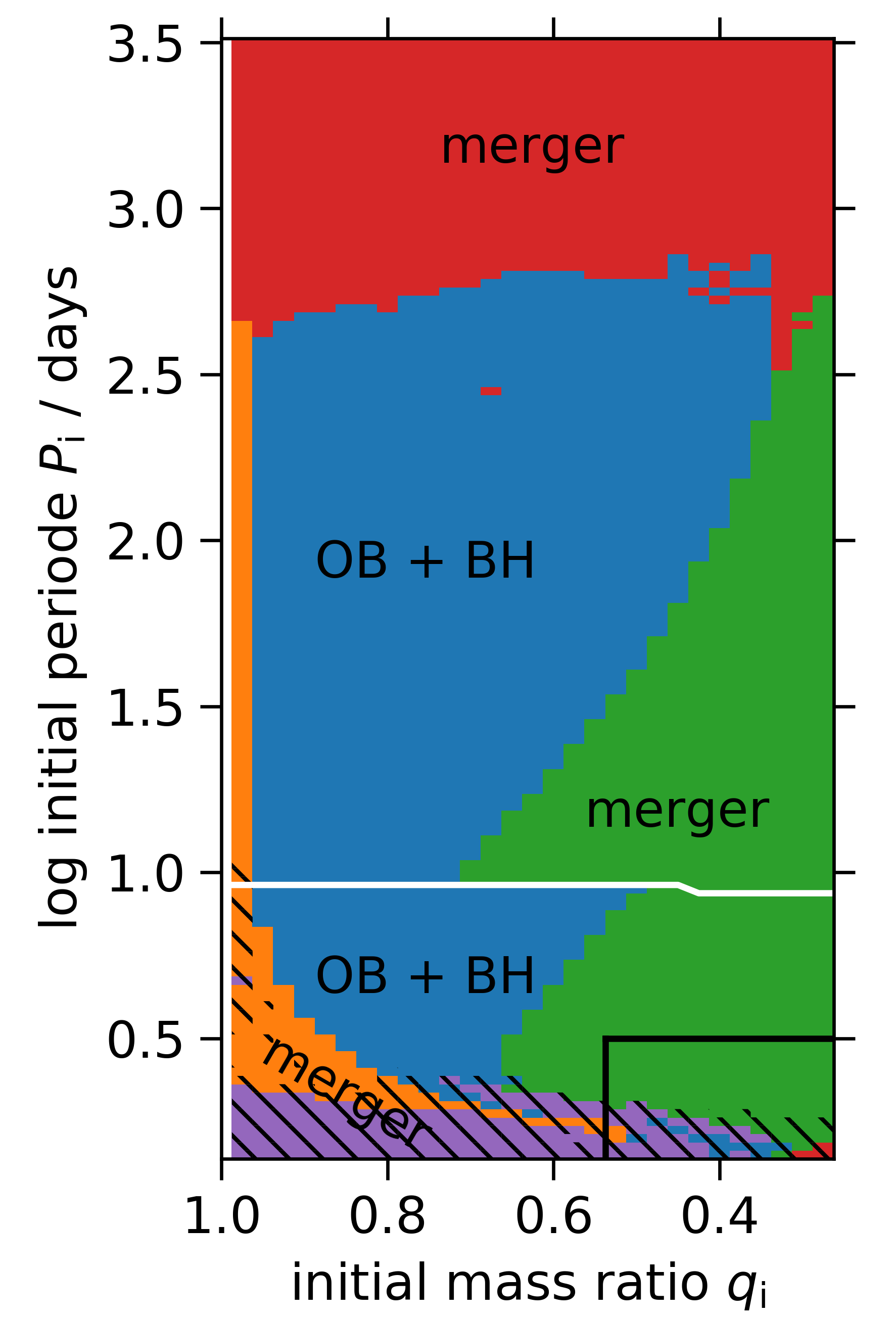}
   \caption{Outcome of the 4020 binary evolution models with an initial primary mass
   of $\log M/\mso = 1.4$ ($\sim 25.12\mso$), as function of their initial orbital period $P_{\rm i}$ and mass ratio
   $q_{\rm i}$. Each of the $30 \times 134$ pixel in this plot represents one detailed binary evolution model.
	   The dark blue systems evolve to the OB+BH stage. {Systems that evolve into a contact configuration are marked by
   black hatching. Purple color indicates systems evolving into mass overflow at the outer Lagrangian point L2,
   and systems evolving into inverse mass transfer occuring from a post main sequence component 
   are marked in yellow; we assume the binaries merge in both situations.
   We also assume those systems to merge which exceed the upper mass-loss rate limit (see main text), marked in green.	   
   The systems with the largest initial orbital periods, marked in red, impart a classical common envelope evolution;
   for simplicity, we assume all of them to merge as well.
%  Hatching indicates that a strong wind from the accretor is required to avoid merging (see text). 
	   Systems below the nearly horizontal white line }
   undergo the first mass transfer while both stars are core hydrogen burning (Case\,A), while the primaries in
   initially wider systems starts mass transfer after core hydrogen exhaustion (Case\,B). 
   The area framed by the black line in the lower right corner marks the part of the parameter space
   which is disregarded in our results (see Sect.\,\ref{sec:method}). Equivalent plots for four more initial primary masses
   are provided in the Appendix.}
   \label{fig:grid25}
   \end{figure}

An inspection of the detailed results showed that some of the contact systems were erroneous. 
In these cases, the primary kept expanding after contact was reached, but no mass transfer was computed. 
This situation is unphysical. An example case is the model with the initial parameters 
$(\log M_{\rm 1,i}, q_{\rm i}, \log P_{\rm orb,i}) = (1.4, 0.4, 0.2)$. 
In Fig.\,\ref{fig:grid25}, this concerns the 10 blue pixel inside the frame in the lower
right corner. The error lead to these systems surviving until, including, the OB+BH stage.  
The error did not occur for initial mass ratios above 0.5. In a recalculation of several of the 
erroneous systems with MESA Version No.\,12115, the unphysical situation did not occur. In these calculations,
the systems merged while both stars underwent core hydrogen burning. In order to avoid any
feature of the erroneous models in our results for OB+BH binaries, we disregard by hand binary models
for which simultaneously $q_{\rm i}<0.55$ and $\log P_{\rm orb,i} < 0.5$, such that none of the non-erroneous systems
in this part of the parameter space contributes to the OB+BH binary population. 
{These systems remain to be considered duing their pre-interaction evolution.}

To account for OB+BH systems, we assess the helium core masses of our models. We consider the pre-collapse single star models of
of Sukhbold et al. (2018), who evaluate the explodability of their models based on their so-called compactness parameter
(O'Connor \& Ott 2011, Ugliano et al. 2012).
Near an initial mass of 20$\mso$, this parameter shows a sudden increase, with most stellar models below this
mass providing supernovae and neutron stars, and most models above this mass expected to form BHs. This mass threshold
is essentially confirmed by Ertl et al. (2016) and M\"uller et al. (2016) based on different criteria, and corresponds
to a final helium core mass of 6.6$\mso$, and a final CO-core mass of 5$\mso$ (Sukhbold et al. 2018). 
Sukhbold et al. also find the threshold to depend only weakly on metallicity. Whereas these
three papers all predict a non-monotonous behaviour as function of the initial mass, with the possibility of
some successful supernovae occurring above 20$\mso$, we neglect this possibility for simplicity, and assume BHs to form
in models with a helium core mass above 6.6$\mso$ at the time of core carbon exhaustion. 

While our adopted BH formation criterion is based on single stars, it has been argued that in stripped stars, the
helium core does not grow in mass during helium burning, such that the $^{12}C$-abundance remains larger,
which ultimately leads to a higher likelihood for NS production than in corresponding single stars (Brown et al. 2001).
On the other hand, recent pre-collapse models evolved from helium stars (Woosley 2019) show a similar 
jump of the compactness parameter as quoted above. The onset of this jump is shifted to larger helium core masses
by about 0.5$\mso$, while the peak is shifted by $\sim 2\mso$. Also, the helium star models predict an island
of low compactness in the He-core mass range $10\dots 12\mso$ which is absent or much reduced in the models which are 
clothed with a H-rich envelope. 
Therefore, with our BH formation criterion as mentioned above, we may overpredict relatively low-mass black holes.
We discuss the corresponding uncertainty in Sect.\ref{sec:uncertain}.

We further assume that the mass of the BH is the same as the mass of the He-core of its progenitor, and that the BHs
form  without a momentum kick. The validity of these assumptions depends on the amount of neutrino energy injection into
the fall-back material after core bounce (Batta et al. 2017). In the direct collapse scenario, the BH forms
very quickly, and a strong kick and mass ejection from the helium star may be avoided. However, in particular near
the NS/BH formation boundary, both assumptions may be violated to some extent. This introduces some additional
uncertainty for our model predictions in the lower part of the BH mass range (cf., Sect.\ref{sec:uncertain})

Due to the high density of our binary evolution grid, it is well suited to construct synthetic stellar populations.
In order to do so, sets of random initial binary parameters are defined under the condition that they obey chosen initial distribution
functions. This is done here by requiring that the primary masses follow the Salpeter (1955) initial mass function,
and that the initial mass ratios and orbital periods follow the distributions obtained by Sana et al. (2013, see also Almeida et al. 2017)
for the massive stars observed in the VLT FLAMES Tarantula survey (Evans et al. 2011). 
The adopted IMF should serve to
constrain the lower limits on the number of systems (cf. adopting the shallower value for 
the 30 Doradus region from Schneider et al. 2018).
%The physical properties
%of the binaries with the selected initial parameters are then linearly interpolated in our binary model grid.

One may select models at a predefined age to construct synthetic star clusters (cf., Wang et al. 2020),
or, as done here, consider a constant star formation rate. We then consider a given
binary model an OB+BH system if it fulfills our BH formation criterion for the 
initially more massive star, and if the initially less massive star is still undergoing core hydrogen burning 
($X_{\rm c} \geq 0.01$). We then consider its statistical weight in accordance with the above mentioned 
distribution functions, and its lifetime as OB+BH binary. With this taken into account, its properties are evaluated at the time of BH formation.

%______________________

\section{Results}
\label{sec:results}
%______________________________________________

   \begin{figure}
   \centering
   \includegraphics[width = \linewidth]{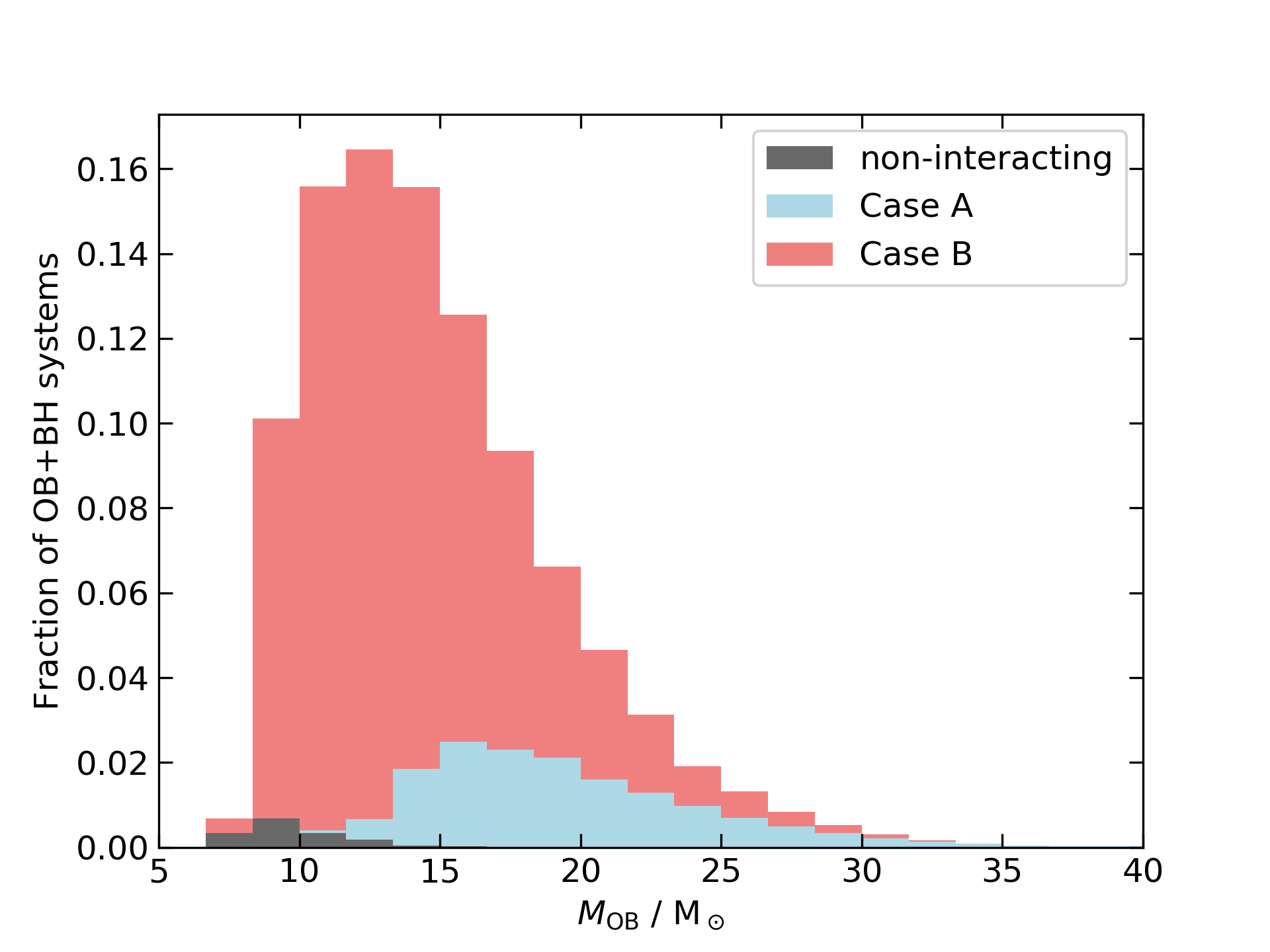}
   \includegraphics[width = \linewidth]{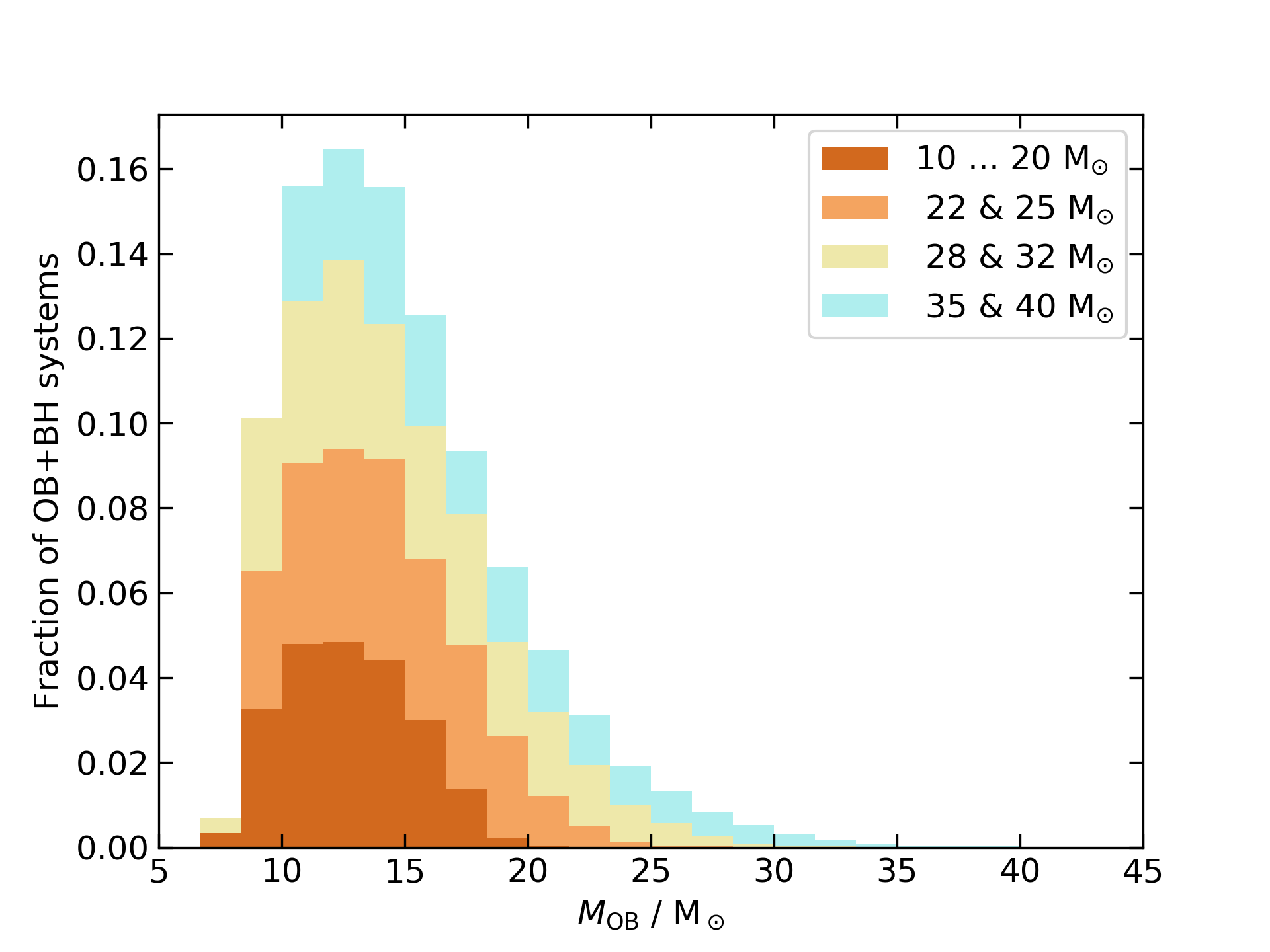}
   \caption{Top: Distribution of the OB star masses of systems in our
   binary evolution model grid that reach the OB+BH stage, assuming constant star formation, weighted with the
   IMF and the initial binary parameter distribution functions, and with their 
   lifetime as OB+BH binary. The red and blue areas represent Case\,B and Case\,A
   systems. Black colour indicates the small number of non-interacting
   systems in our binary grid. The results are stacked, such that the upper envelope 
   corresponds to the total number of systems. 
   The ordinate values are normalised such that the value
   for each bin gives its relative contribution to the total number of systems.
   Bottom: the same distribution as in the top plot, but discriminating between
   different initial masses of the BH prigenitors (see legend).}
   \label{fig:m}
   \end{figure}
%______________________________________________

As we focus on the properties of OB+BH binaries in this paper, in the following
we discuss only systems that avoid to merge before they form the first compact object.
In that, it is useful to consider the Case\,A systems separately from the Case\,B systems. Not only
are the predictions from both classes of binaries quite distinct from each other (see below), but 
also the physics which is involved in the mass transfer process.

To a large extent, tidal effects can be neglected in the wider Case\,B systems, while they play an important role
in Case\,A systems. In the latter, tidal coupling slows down
or prevents the spin-up of the mass gainer during mass transfer, while direct impact accretion also reduces the specific angular
momentum of the accreted matter (Langer 2012). Consequently, the mass transfer efficiency, i.e., the ratio of
the mass accreted by the mass gainer over the amount of transferred mass, can be high in Case\,A systems.
We find accretion efficiencies of up to nearly one,
with an average of about 30\% for all Case\,A binaries, and the highest values are achieved for the most massive systems and largest
initial mass ratios (i.e., $q\simeq 1$). In contrast,
the mass transfer is rather inefficient in most of our Case\,B systems, because the
mass gainer is quickly spun-up to critical rotation, such that any further accretion remains very limited.
The overall accretion efficiency remains at the level of 10\% or less. 

\subsection{OB star masses, BH masses, mass ratios}

%______________________________________________

   \begin{figure}
   \centering
   \includegraphics[width = \linewidth]{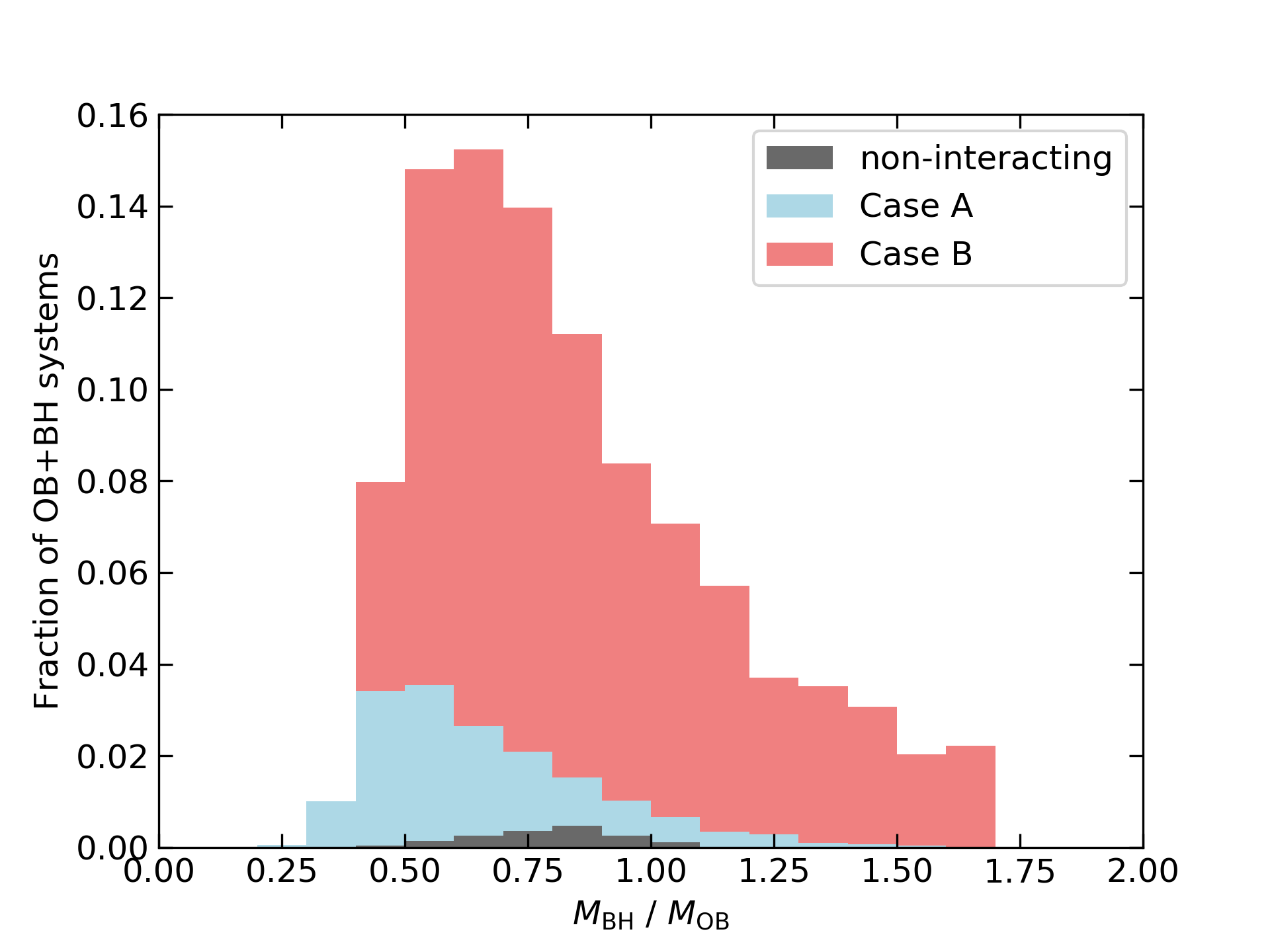}
   \includegraphics[width = \linewidth]{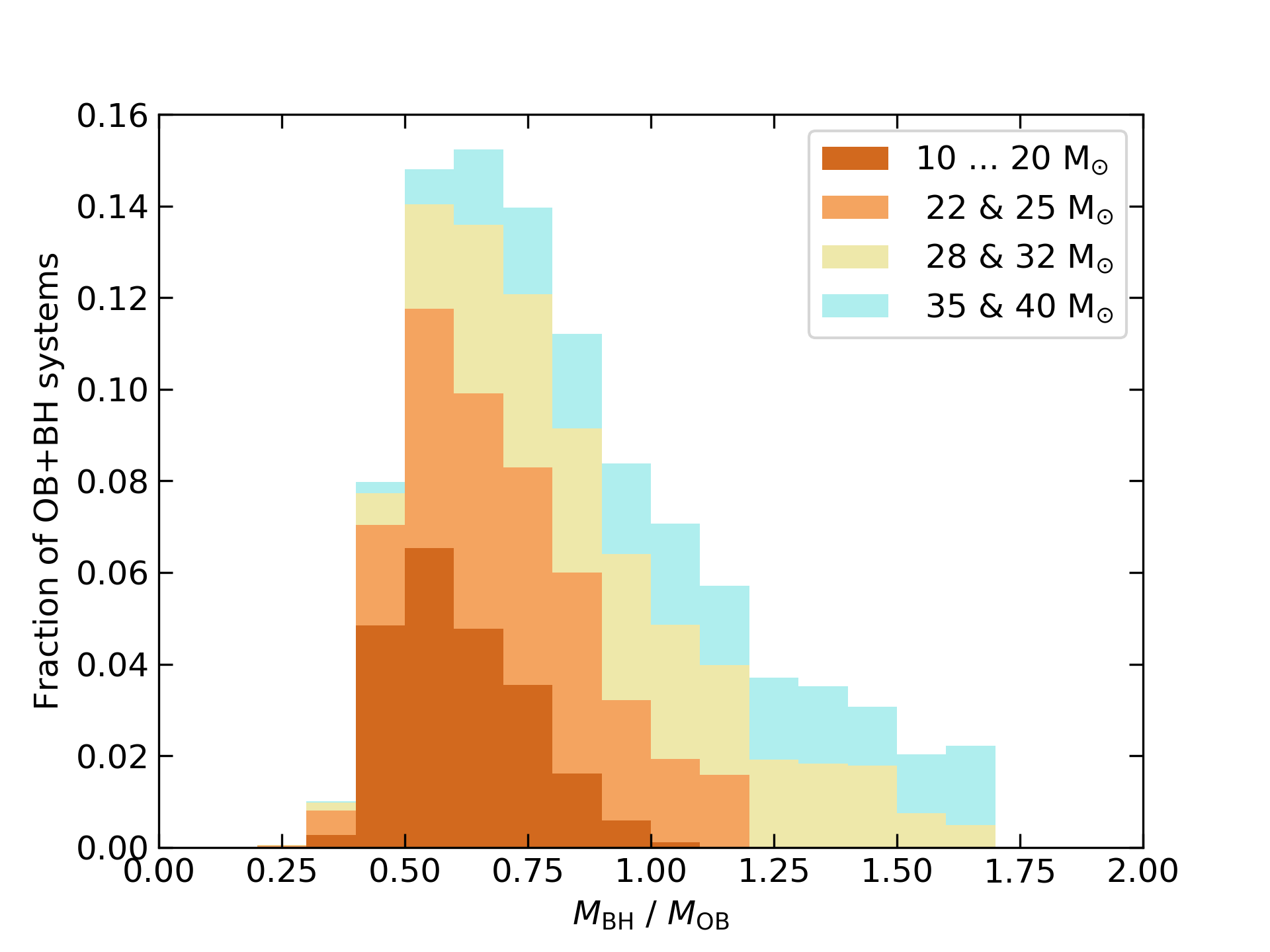}
   \caption{Top: As Fig.\,3, here showing the distribution of the BH/OB-star mass ratios 
   in our predicted OB+BH binaries.
   Bottom: the same distribution as in the top plot, but discriminating between
   different initial masses of the BH progenitors (see legend).}
   \label{fig:q}
   \end{figure}
%______________________________________________

As found in previous binary evolution calculations (e.g., Yoon et al., 2010), 
the mass donors of our model binaries are stripped of nearly their entire hydrogen envelope
as consequence of Roche-lobe overflow. Whereas small amounts of hydrogen may remain in the
lower-mass primaries (Gilkis et al., 2019), it is reasonable to consider them as helium stars
after the mass transfer phase.
Whereas the initial helium star mass emerging from Case\,B binaries is very similar to the initial helium core mass 
(i.e., at core helium ignition) of single stars, we emphasise that, {due to larger amounts of mass being transferd 
during the MS stage}, Case\,A binaries produce significantly smaller mass
helium stars (cf., fig.\,14 of Wellstein et al., 2001), an effect which is mostly not accounted for in 
simplified binary evolution models. 

Figure\,\ref{fig:m} evaluates the distribution of the masses of the OB\,stars in our OB+BH models 
at the time of the formation of the first compact object. Besides the Case\,A and\,B systems, 
it distinguishes for completeness, the systems in our
grid which never interact. The results shown in Fig.\,\ref{fig:m} are weighted by the initial mass
and binary parameter distribution functions (see Sect.\,2), and by the duration of the OB+BH phase of the individual
binary models. As such, Fig.\,\ref{fig:m} predicts the measured distribution of the OB\,star masses in 
idealised and unbiased observations of OB+BH binaries. 

The distribution of the masses of the OB\,stars in our OB+BH binaries shown in 
Fig.\,\ref{fig:m} has a peak near $14\mso$. 
Towards lower OB masses, the chance that the final helium core mass of the mass donor falls below our
threshold mass for BH formation is increasing. Whereas for the donor star initial masses, there is a cut-off near
$18\mso$ below which no BHs are produced, the distribution of the masses of their companions leads
to a spread of the lower mass threshold of the secondaries, i.e., the OB stars in BH+OB systems,
leading to the lowest masses of the BH companions of about $8\mso$. 
The drop of the number of systems for OB star masses above $14\mso$ is mainly produced by
the initial mass function, and by the shorter lifetime of more massive OB stars.
Due to the limits of our model grid to initial primary masses below $40\mso$, we may
be missing stars in the distribution shown in Fig.\,\ref{fig:m} above $\sim 20\mso$.
However, their contribution is expected to be small, and it is very uncertain since the
corresponding stars show envelope inflation (cf., Sect.\,\ref{sec:uncertain}). 

The upper panel of Fig.\,\ref{fig:m} shows that the majority of OB+BH systems is produced via
Case\,B evolution, as expected from Fig.\,\ref{fig:grid25} when the areas covered by Case\,A and Case\,B
in the $q_{\rm i}-P_{\rm i}$-plane are compared {(though note that our initial distributions
are not exactly flat in $\log P_{\rm i}$ and $q_{\rm i}$)}. 
The peak in the OB mass 
distribution of the Case\,A models is shifted to higher masses ($\sim 16\mso$), compared to the Case\,B distribution
due to the higher accretion efficiency in Case\,A. 
For the same reason, the most massive OB stars in the OB+BH systems produced by our grid, with
masses of up to $47\mso$, evolved via Case\,A (cf., Sect.\,\ref{sec:detect}). 
The Case\,B binaries produce only OB\,star companions to BHs with masses below $\sim 34\mso$, 
notably since the most massive Case\,B systems with mass ratios above $\sim 0.9$ lead to mergers before the BH is formed. 

The bottom panel of Fig.\,\ref{fig:m} provides some insight into the mass dependence of the production
of OB+BH binaries (see also bottom panel of Fig.\,\ref{fig:q}),
by comparing the contributions from binary systems with four different initial primary
mass ranges. We see that systems with successively more massive primaries
produce more massive OB stars in OB+BH binaries. We also see that the range of OB star masses
in OB+BH binaries originating from systems with more massive primaries is larger.
This reflects our criterion for mergers in Case\,B systems (Sect.\,\ref{sec:method}),
which implies that it is easier for more massive binaries to drive the excess mass which the spun-up mass
gainer can not accrete any more out of the system. 
%Finally, we see that whereas systems with
%donor masses of up to 20$\mso$ can produce OB+BH binaries with $\sim 20\mso$ OB\,stars, 
%systems with initial donor masses of 35$\mso$ donor stars produce mostly OB\,stars in OB+BH binaries with masses below $35\mso$,
%due to stellar wind mass loss. 

Figure\,\ref{fig:q} shows the resulting distribution of mass ratios of our OB+BH binary models,
produced with the same assumptions as Fig.\,\ref{fig:m}. Remarkably, the distribution drops
sharply for BH/OB star mass ratios below 0.5. The main reason is that the BH is produced by
the initially more massive star in the binary. This means that binaries with a small initial mass ratio
(say $M_{\rm 2,i}/M_{\rm 1,i}\simeq 1/3$; cf., Fig.\,\ref{fig:grid25}) easily produce 
black holes as massive as their companion or more, such that their BH/OB mass ratios
is one or higher. Since the accretion efficiency in our models is mostly quite low,
binaries starting with a mass ratio near one, on the other hand, obtain BH/OB mass ratios
larger than 0.3 since more than one third of the primaries' initial mass ends up in the BH.
Since the corresponding fraction is larger in more massive primaries, we find 
that more massive primaries lead to larger BH/OB mass ratios, where those with
initial primary masses below 20$\mso$ produce only OB+BH binaries with $M_{\rm BH}/M_{\rm OB} < 1$
(Fig.\,\ref{fig:q}, bottom panel).

%______________________________________________
   \begin{figure}
   \centering
   \includegraphics[width = \linewidth]{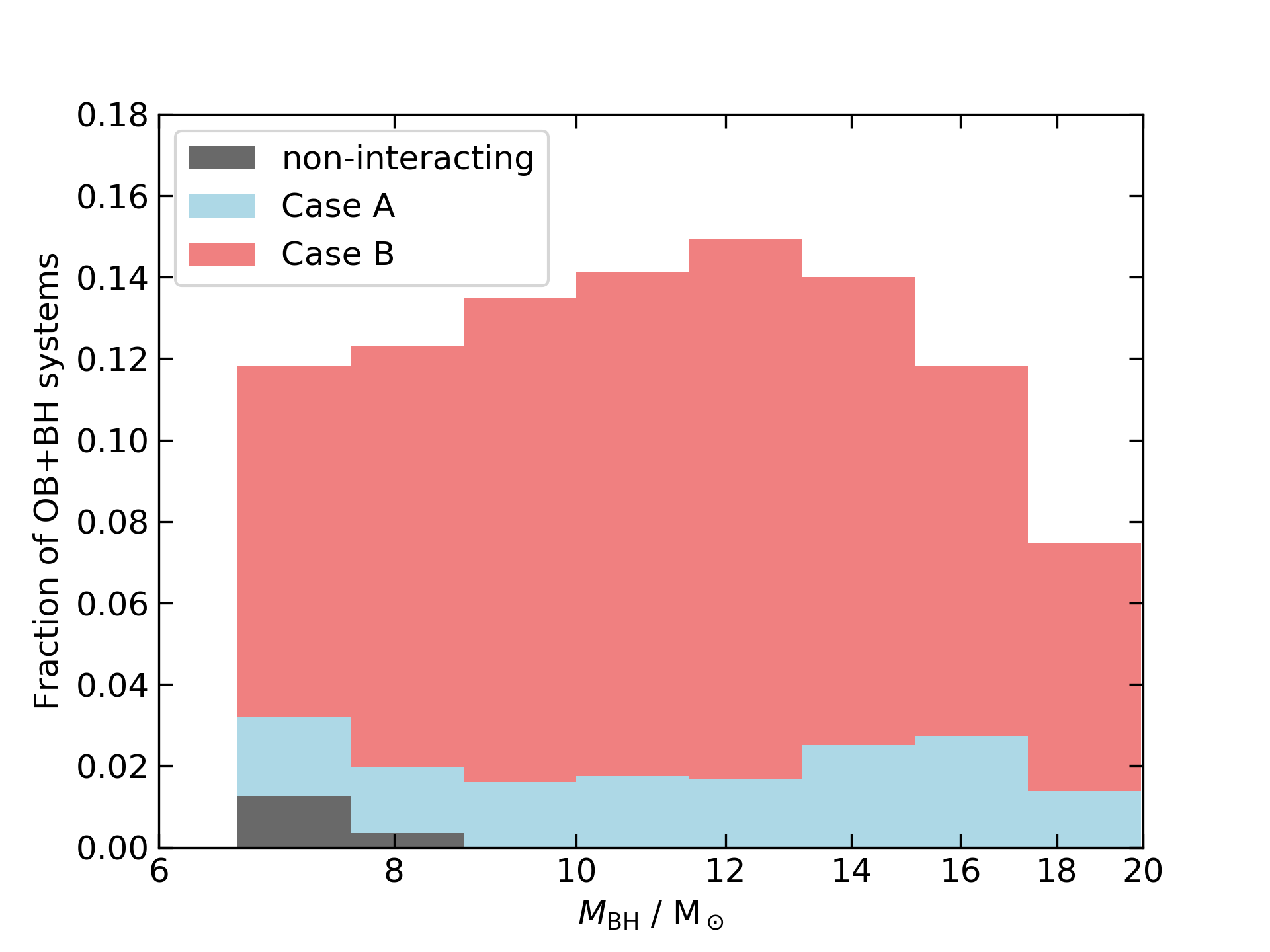}
   \caption{As Fig.\,3, here showing the distribution of the BH masses
   at the time of BH formation in our predicted OB+BH binaries.}
   \label{fig:mbh}
   \end{figure}
%______________________

The distribution of the BH masses produced in our binaries shows a broad peak near 10$\mso$ (Fig.\,\ref{fig:mbh}),
with a sharp lower limit of $6.6\mso$ as introduced by our assumptions on BH formation (Sect.\,2).
While the drop in the IMF towards higher masses leads to a decrease of the number of BHs for increasing
BH mass, this effect is less drastic than for the OB star mass (Fig.\,\ref{fig:m}). This can be understood
by considering the systems with the most massive primaries in our grid, which form the most massive BHs.
These systems produce OB+BH binaries with a broad range of OB star masses (blue part in bottom panel of
Fig.\,\ref{fig:m}), such that their contribution to Fig.\,\ref{fig:mbh} will benefit from a broad range
of durations of the OB+BH phase. 
The masses of the produced BHs in our grid are limited to about $22\mso$, in agreement with
earlier predictions (Belczynski et al. 2010). This is due to the heavy wind mass loss
of the BH progenitors during their phase as Wolf-Rayet star and may therefore be strongly metallicity dependent.

\subsection{Orbital periods and velocities}
%______________________________________________
   \begin{figure}
   \centering
   \includegraphics[width = \linewidth]{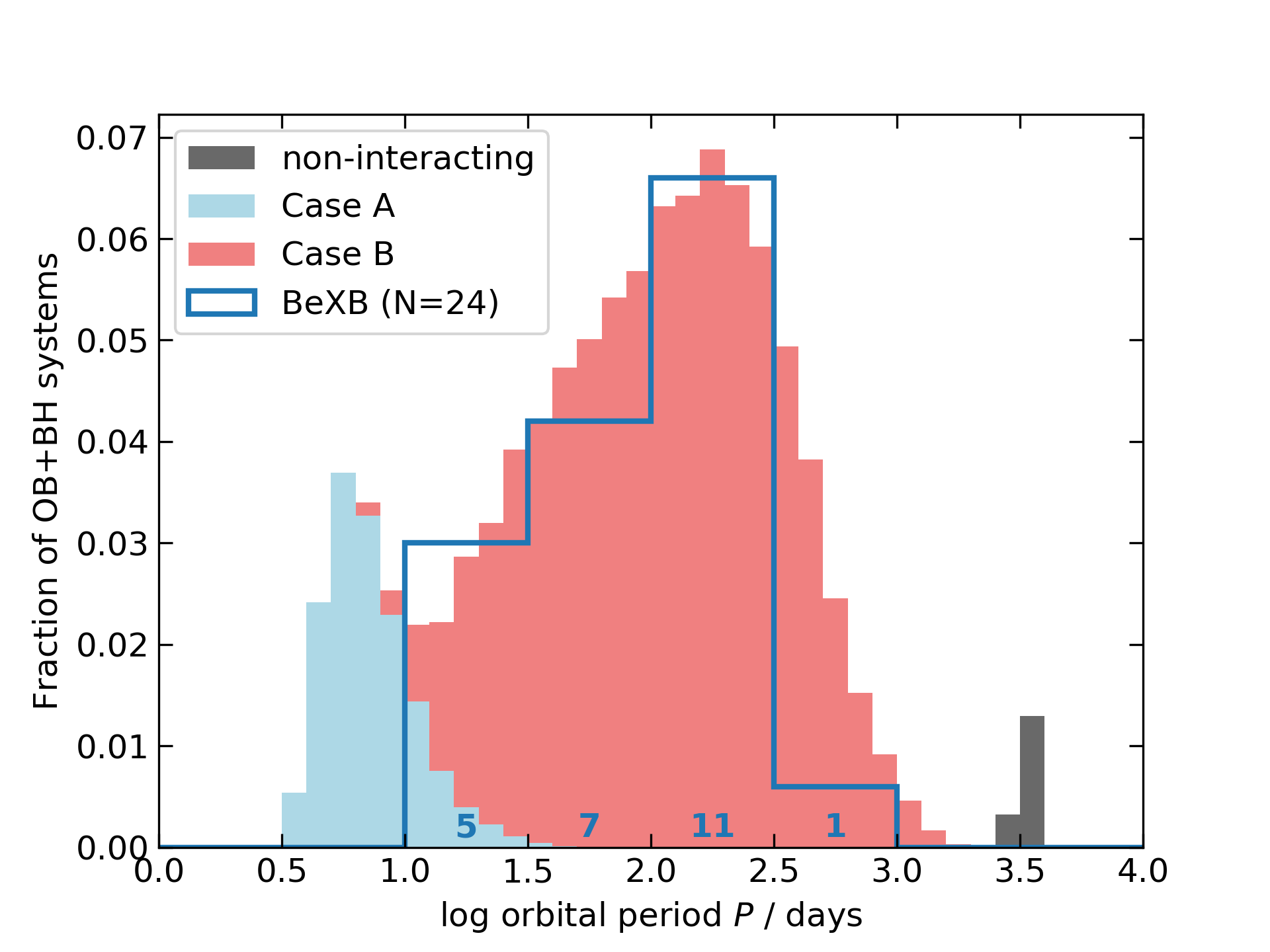}
   \includegraphics[width = \linewidth]{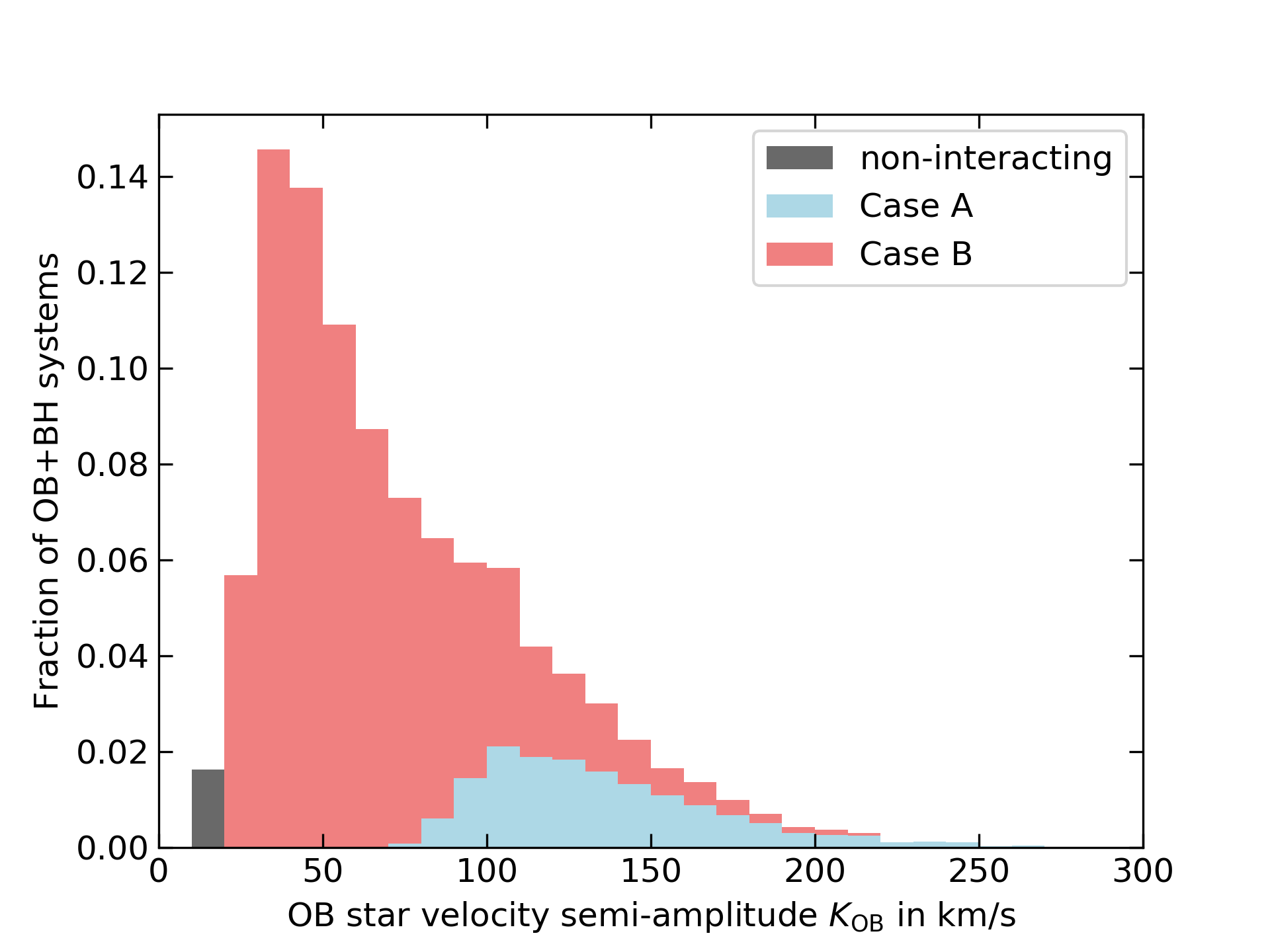}
   \caption{As Fig.\,3, here showing the distribution of the orbital periods at the 
   time of BH formation (top), and
   of the orbital velocity amplitudes (bottom) of our OB+BH binaries.
   The blue line in the top plot shows the distribution of the orbital periods
   of the Galactic Be/X-ray binaries (Walter et al. 2015).}
   \label{fig:p}
   \end{figure}
%______________________
%______________________

   \begin{figure}
   \centering
   \includegraphics[width = \linewidth]{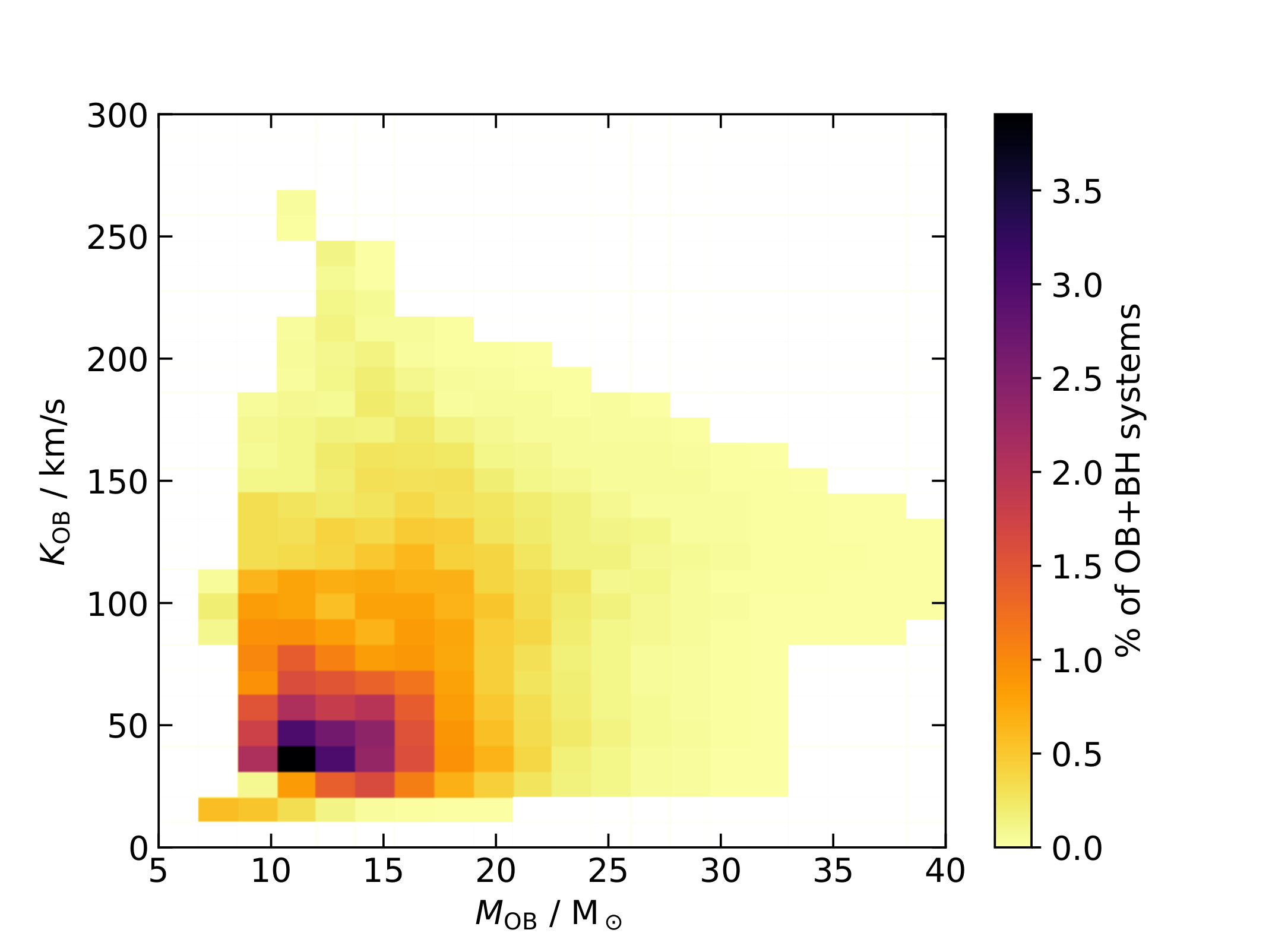}
   \includegraphics[width = \linewidth]{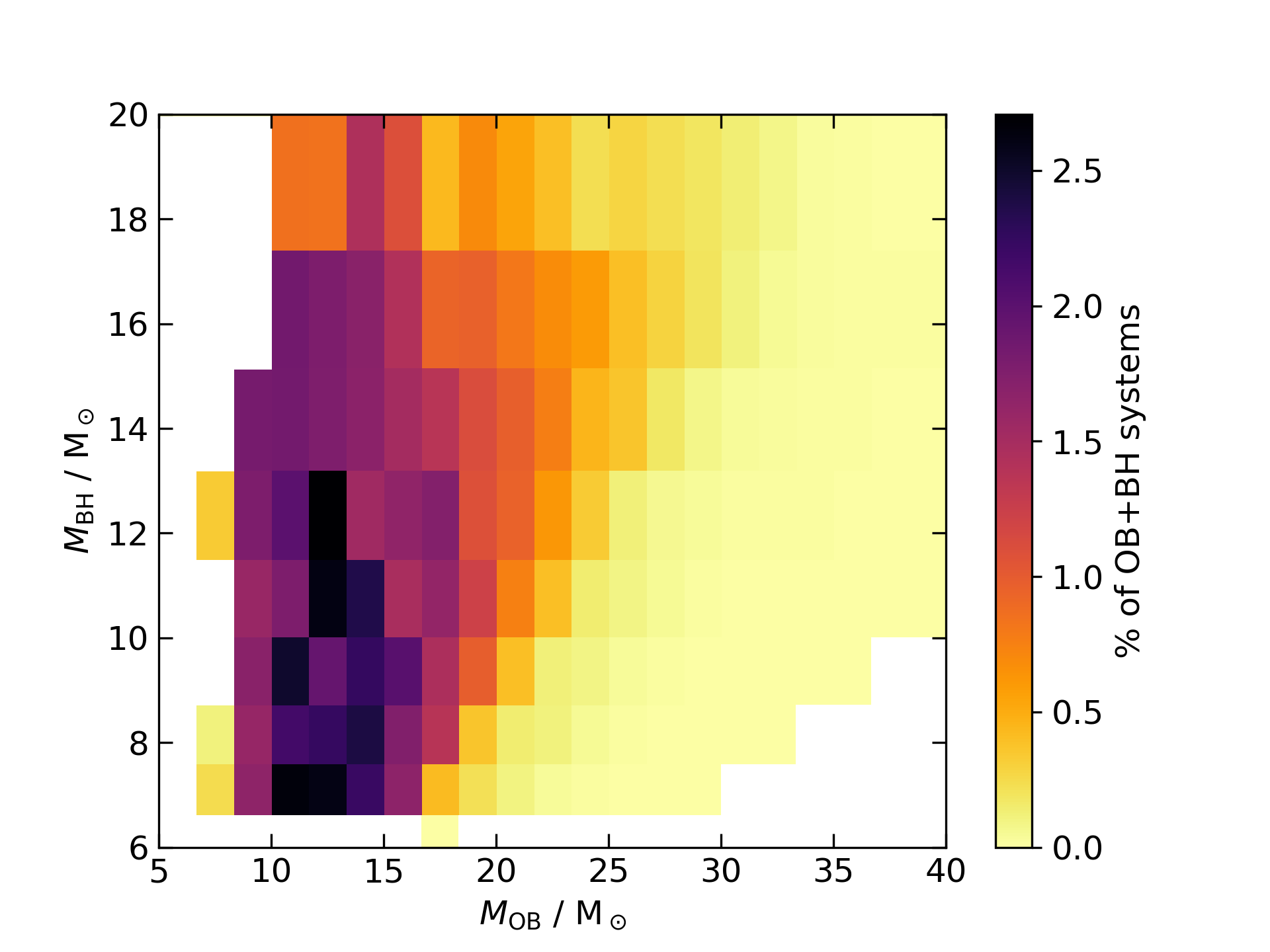}
   \caption{Predicted number distribution of OB+BH systems in the parameter space
OB star mass$-$orbital velocity (top panel) and OB star mass$-$black hole mass
(bottom panel). The expected numbers in each pixel are colour-coded and normalised such that
the sum over all pixels is 100\%.}
   \label{fig:2d}
   \end{figure}
%______________________
The top panel of Fig.\,\ref{fig:p} shows the predicted distribution of orbital periods
of the OB-BH binaries found in our model grid. We find that non-interacting binaries
may produce OB+BH binaries with orbital periods in excess of about 3\,yr. In Fig.\,\ref{fig:p}
we can show only the shortest-period ones due to the upper initial period bound of our binary grid.
Perhaps, many more such binaries may form, even small black hole formation kicks could break them
up, the easier the larger the period. As these systems would also be the hardest to observed, we focus here on
OB-BH binaries which emerge after mass transfer due to Roche-lobe overflow. 

As seen in Fig.\,\ref{fig:p}, the distribution of these post-interaction OB+BH binaries shows two
distinct peaks, which we can attribute to the two distinguished modes of mass transfer.
Not surprisingly, the Case\,A systems are found at smaller periods and remain below $\sim 30\,$d, 
while the Case\,B systems spread between about 10\,d and 1000\,d, with a pronounced maximum
near 150\,d. 
Overplotted in Fig.\,\ref{fig:p} is the observed orbital period distribution of 24\,Galactic 
Be/X-ray binaries. We discuss the striking similarity with the period distribution of our OB+BH models
in Sect.\,\ref{sec:comobs}).

Through Kepler's laws, we can convert the period distribution into a distribution of orbital velocities
of the OB star components in OB+BH systems, which we show in the bottom panel of Fig.\,\ref{fig:p}. 
As expected, the orbital velocities are largest in Case\,A binaries, and smallest in the Case\,B systems. 
These values are all so large that they are easily measurable spectroscopically
(cf., Sect.\,\ref{sec:detect}). 

{Figure\,\ref{fig:2d} illustrates the 2D-distributions of both component masses and the orbital velocity.
In accordance with Fig.\,3, we see that the OB masses are strongly concentrated to the mass range
$8\mso\dots 25\mso$. The top figure shows that the OB+BH binaries are most abundant in
a small area of the orbital velocity vs. OB mass plane, i.e., near $M_{\rm OB}\simeq 13\mso$ and
$K_{\rm OB}\simeq 50\,$km/s. More than half of all systems are expected to have OB masses below $17\mso$
with orbital velocities of $K_{\rm OB} < 70\,$km/s. At the same time,
the bottom plot of Fig.\,\ref{fig:2d} shows that the expected BH companions to $\simeq 13\mso$ 
B\,stars have a rather flat distribution between $7\mso$ and $20\mso$ (see also Fig.\,5).}

\subsection{OB star rotation and surface abundances}

%______________________________________________

   \begin{figure}
   \centering
   \includegraphics[width = \linewidth]{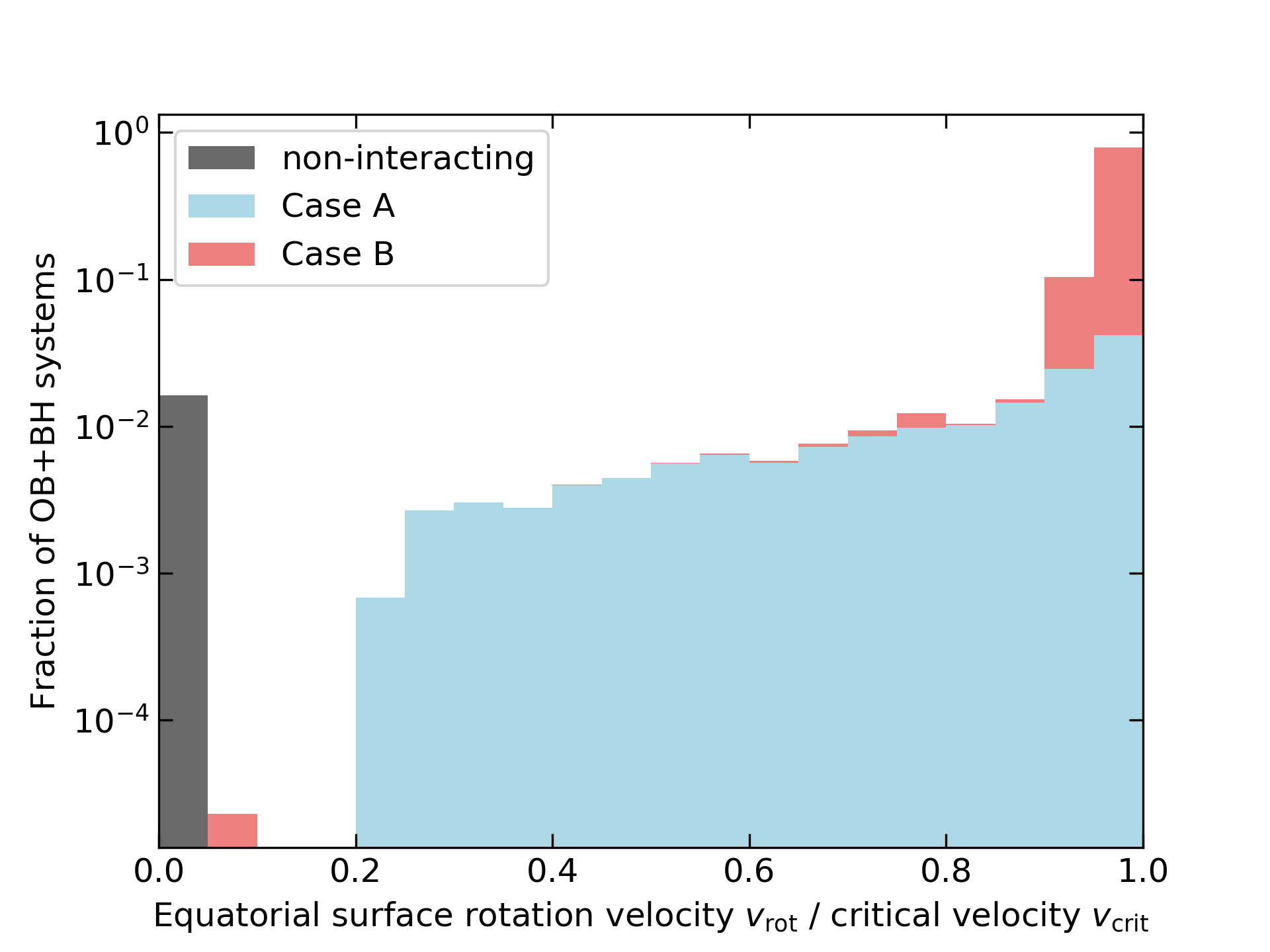}
   \includegraphics[width = \linewidth]{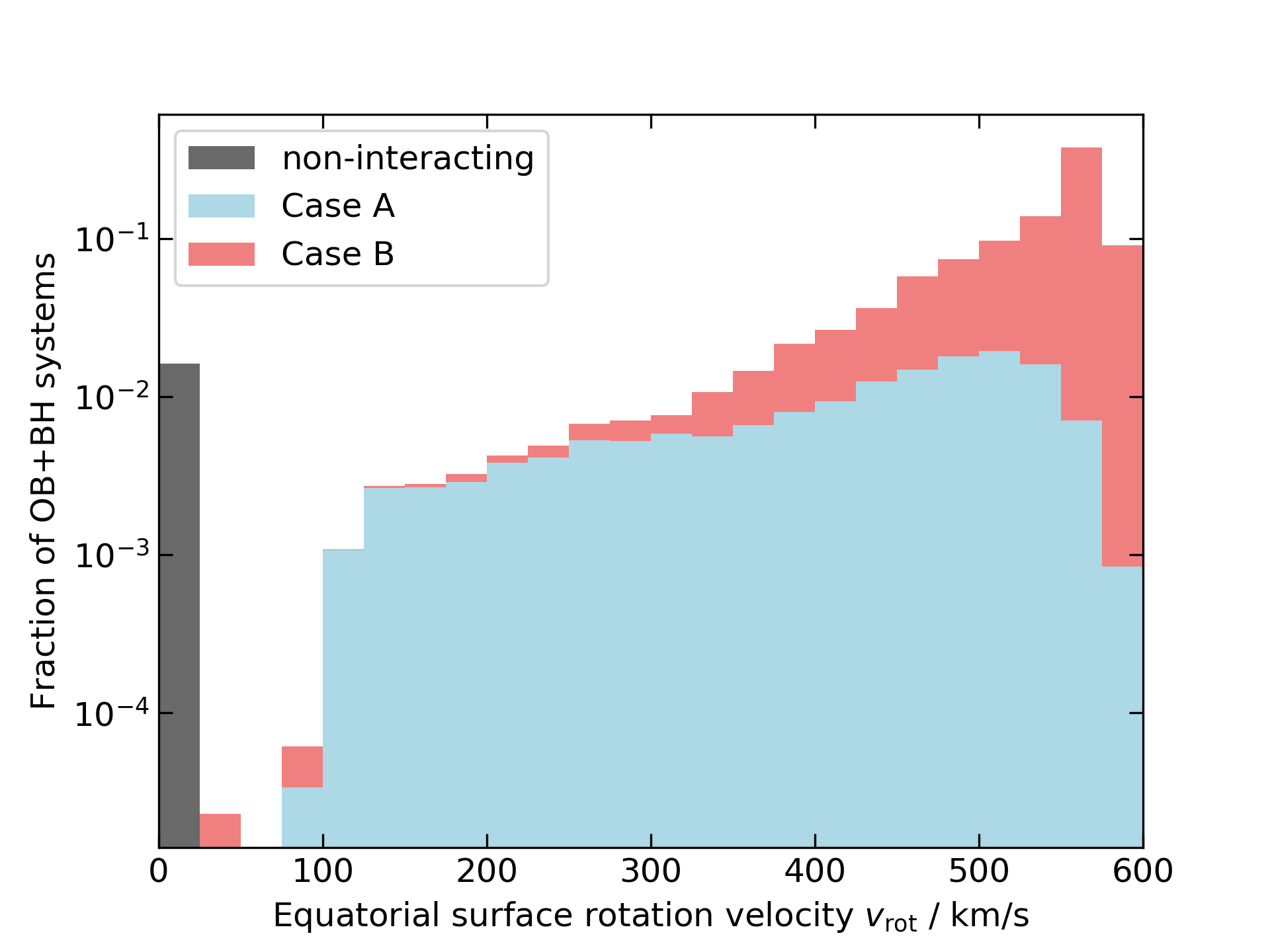}
   \caption{Distribution of the ratio of equatorial surface rotation velocity
   to critical rotation velocity for the OB stars in
   OB+BH binaries, at the moment of BH formation, as predicted by our population synthesis model (top panel).
   The bottom panel shows the corresponding distribution of the absolute equatorial surface rotation velocities
   of the OB\,stars as obtained in the indicated mass bins. In both plots, the small peak near zero
   rotation is due to the widest, non-interacting binaries; it is non-physical and should be disregarded.}
   \label{fig:rot}
   \end{figure}
%______________________

%______________________________________________

   \begin{figure}
   \centering
   \includegraphics[width = \linewidth]{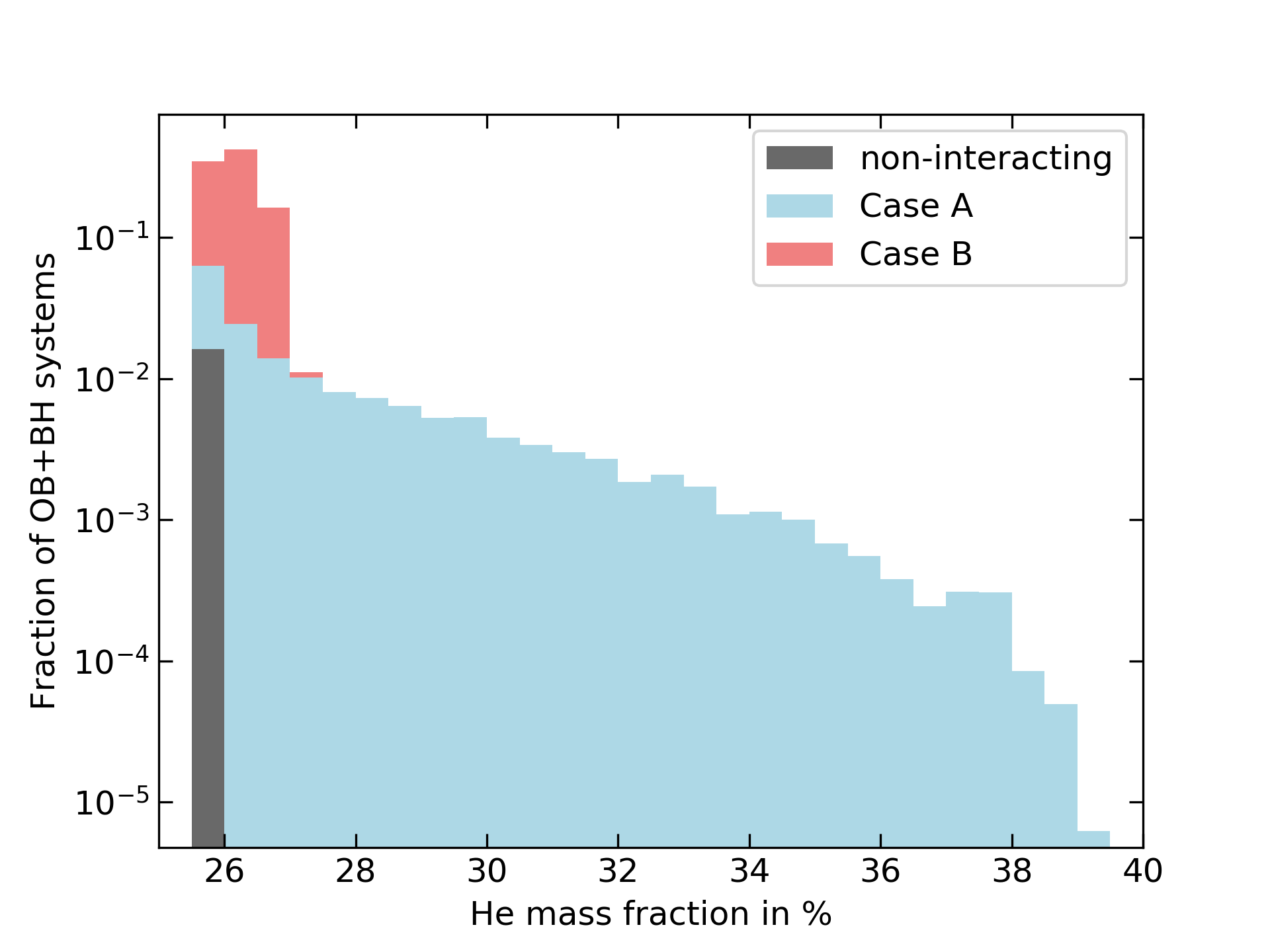}
   \includegraphics[width = \linewidth]{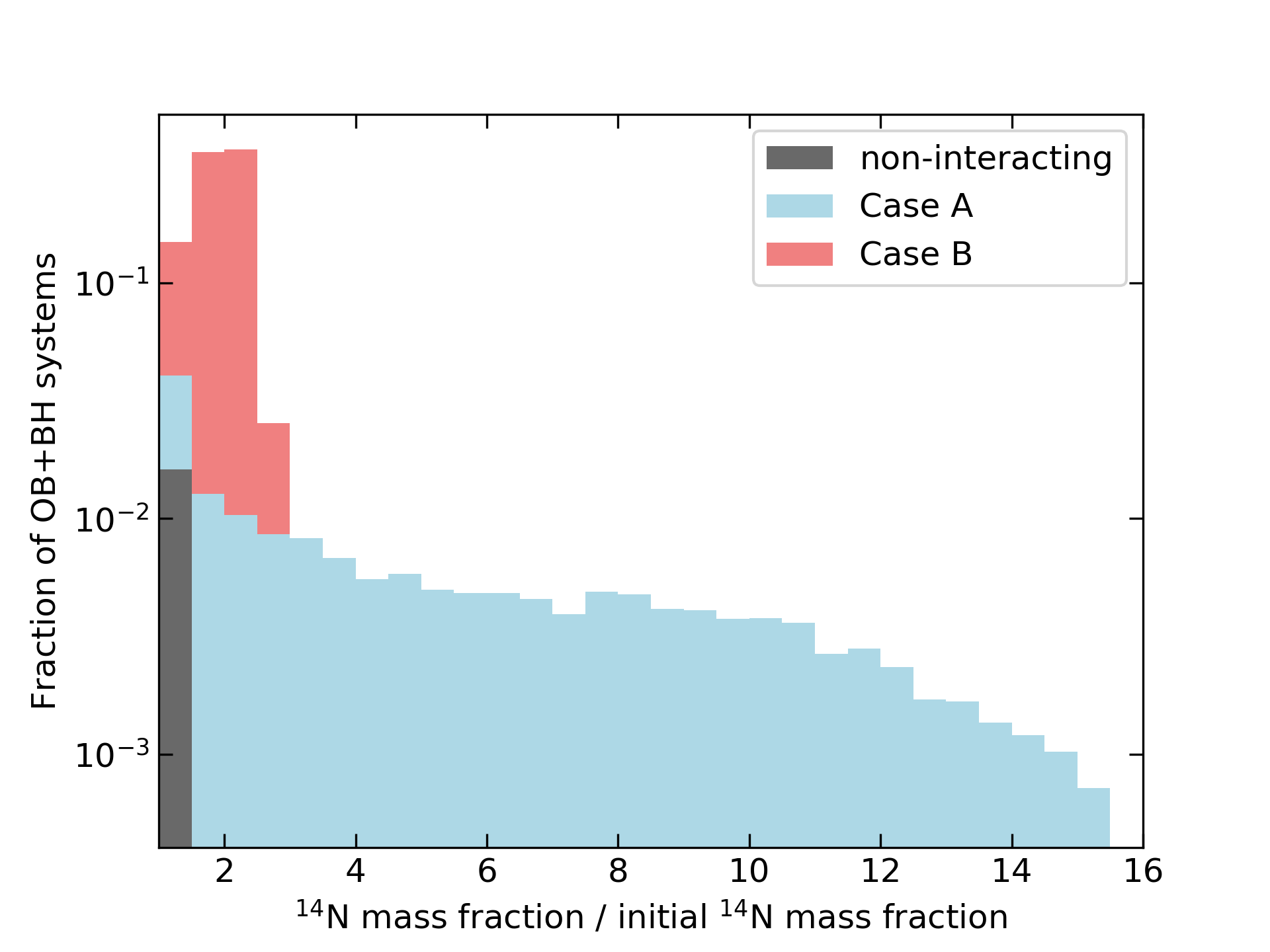}
   \caption{Result of our population synthesis calculations for the probability distribution
   of the surface helium (top) and nitrogen (bottom) surface abundances of the OB stars in
   OB+BH binaries.}
   \label{fig:yn}
   \end{figure}
%______________________

As pointed out in Sect.\,\ref{sec:method}, our detailed binary stellar evolution models 
accurately keep track of the angular momentum budget of both stars. They consider 
internal angular momentum transfer due to differential rotation, angular momentum loss by winds,
angular momentum gain by accretion, and spin-orbit angular momentum exchange due to tides.  

Figure\,\ref{fig:rot} shows that most of the OB components in our OB+BH binary models are
rapid rotators. At the time of BH formation, as many as half of them rotate very close to critical
rotation. In particular, a high fraction of those systems which originate from Case\,B mass transfer, 
where tidal breaking is unimportant, rotate very close to critical.
The Case\,A systems have a much broader distribution in Fig.\,\ref{fig:rot}. The minimum value
of $\varv_{\rm rot}/\varv_{\rm crit}=0.2$ corresponds to the widest systems where tidal breaking still works,
i.e., where the synchronisation timescale becomes comparable to the nuclear timescale of the OB star.

Looking at the absolute values of the rotational velocities, the bottom panel of Fig.\,\ref{fig:rot} reveals
a broader distribution. This is mostly an effect of the mass and time dependence of the critical rotational
velocity. However, we see that even the Case\,A binaries stretch out to high rotation velocities, such that
on average their rotation rate is much higher than that of an average O\,star (i.e., $\sim 150\kms$,
Ramirez-Agudelo et al., 2013). 

We point out that Fig.\,\ref{fig:rot} depicts the rotation of the OB stars when the BH forms.
In the time span between the end of the mass-transfer induced spin-up process and the BH
formation, which corresponds to the core helium burning time of the BH progenitor in most cases,
the OB star spin may have changed. The same is true for the lifetime of the OB star
with a BH companion. Here, in particular the O\,stars are expected to lose some angular momentum due to their 
(non-magnetic) wind (Langer 1998, Renzo et al. 2017). On the other hand, single B\,stars
are expected to spin-up as a consequence of their core hydrogen burning evolution
(Ekstrom et al. 2008, Brott et al. 2011, Hastings et al. 2020). This explains that the B stars
in our OB+BH binaries (i.e., the OB components with a mass below $\sim 15\mso$), which are
brought to critical rotation due to accretion, remain at critical rotation for their remaining
hydrogen burning lifetime.  

%The bottom panel of Fig.\,\ref{fig:rot} reveals that this spin-down is strongly mass
%dependant. I.e., for the OB components with a mass below $15\mso$, the Case\,A
%($140\kms \simle v_{\rm rot} \simle 280\kms$) and Case\,B ($420\kms \simle v_{\rm rot} \simle 580\kms$)
%distributions are separated by a gap in rotation velocity. In the majority of the Case\,B
%systems, the corresponding B star model rotates almost at critical rotation (note that the
%critical rotation velocity of a 15$\mso$ star decreases from about 600$\kms$ at the ZAMS
%to $\sim 420\kms$ at the TAMS). This implies that no mass loss induced spin-down occurs, and in fact
%correspond single star models increase their $v_{\rm rot}/v_{\rm crit}$-ratio as function of time
%due to internal angular momentum transport (Ekstrom et al. 2008, Brott et al. 2011, Hastings et al.
%2020). The OB components near 20$\mso$ already spin down significantly, as the Case\,B distribution reaches
%down to 320$\kms$. And for more massive OB companions to BHs, we expect few critical rotators,
%but still a far faster rotation than for an average O star.

A second signature of accretion in the OB component of OB+BH binaries may be the presence of
hydrogen burning products at the surface of the OB star. We note that in our models, rotationally induced 
mixing, semiconvection, and thermohaline mixing are included in detail. We find that the main
enrichment effect is produced by the accretion of processed matter from the companion, and the
subsequent dilution through thermohaline mixing. Despite the fast rotation of the OB components,
rotational mixing plays no major role. The reason is that in contrast to rapidly rotating single star
models, the spun-up mass gainers did not have an extreme rotation before the onset of mass transfer.
During that stage, they could establish a steep H/He-gradient in their interior, which provides
an unpenetrable barrier to rotational mixing after accretion and spin-up have happened.

To quantify the obtained enrichment, we show the
distribution of the surface helium and nitrogen abundances of our OB stars with BHs in Fig.\,\ref{fig:yn}.
We can see that the OB stars in Case\,B binaries remain essentially unenriched. The reason for this
is that our Case\,B  mass gainers accrete only small amounts of mass (about 10\% of their initial
mass). Furthermore, this accretion happens early during the mass transfer process, since the accretion
efficiency drops once the stars are spun-up. Therefore, only material from the outer envelope of
the donor star is accreted, which is generally not enriched in hydrogen burning products.
We expect the near-critically rotating OB stars in our Case B systems to be Be stars. 
Given that Be stars are often not or only weakly enriched in nitrogen (Lennon et al. 2005,
Dunstall et al. 2011), in contrast to predictions from rotating single star models,
the population of Be stars may be dominated by binary-interaction products.

In Case\,A binaries, on the other hand, much more mass is accreted, also matter from the deeper
layers of the mass donor, which have been part of the convective core in the earlier stages
of hydrogen burning. As we see, the surface helium mass fraction goes up to $\sim 35$\%.
This is accompanied by a strong nitrogen enhancement by up to a factor of 12.

\section{Key uncertainties}
\label{sec:uncertain}

\subsection{Envelope inflation}

%{\it TT: General comments to Section 4.1: 
%I find that there are too many (incoherent) pieces of text, here in this section and elsewhere related 
%to exactly this important point on inflated envelopes. I suggest to:
%1) start by mentioning that inflated envelopes occur in masses stars and that these envelopes are fully convective.
%2) Hence their mass radius exponent is negative, and thus they expand in response to mass loss.
%3) This is likely to result in a CE evolution.
%4) Whether or not the system survives this CE is still uncertain. The dilute inflated envelope may only create a weak drag force on the companion star, such that the in-spiral might be relatively mild, thus enhancing the chance of survival.
%
%Maybe MESA cannot numerically calculate the outcome of these systems, but please add a few words about what is the expected outcome and why?
%Also, keep in mind to specify (in various parts of the manuscript) that RLO does occur but it is expected to be dynamically unstable.}

The largest considered initial primary mass in the LMC binary evolution model grid of Marchant (2016)
is $39.8\mso$. In a sense, this mass limit is an experimental result, since it was found that for the
next higher initial primary mass to be considered ($44.7\mso$), the MESA code could not compute
through the mass transfer evolution of most systems. This is not surprising, since single star models computed
with very similar physics assumptions (Brott et al. 2011) predict that such stars with LMC
metallicity expand so strongly that they become red supergiants during core hydrogen burning.
From an analysis of the internal structure of these models, Sanyal et al. (2015, 2017) found that this drastic
expansion is a consequence of the corresponding models reaching the Eddington limit in their outer envelopes,
when all opacity sources (i.e., not only electron scattering) are considered in the Eddington limit.

This so-called envelope inflation can be easily prevented to occur in stellar models. The corresponding
envelope layers are convective, and an enhancement of the convective energy transport efficiency
leads to a deflation of the envelope (fig.\,B.1 of Sanyal et al., 2015). However, there is no reason to doubt
the energy transport efficiency of the classical Mixing Length Theory (B\"ohm-Vitense 1958) in this context.
On the contrary, due to the low densities in the inflated envelope, it is evident that vertically moving
convective eddies radiate away their heat surplus faster than they move, implying a small energy transport efficiency
as computed by the standard Mixing Length Theory (Gr\"afener et al. 2012), which is also verified by corresponding
3D-hydrodynamic model calculations (Jiang et al. 2015). The inflation effect has been connected with observations of
so-called Luminous Blue Variables (Gr\"afener et al. 2012, Sanyal et al. 2015, Grassitelli et al. 2020), which are
hydrogen-rich stars; however, inflation is also predicted to occur in hydrogen-free stars (Ishii et al. 1999, Petrovic
et al. 2006, Gr\"afener et al. 2012, Grassitelli et al. 2016).

Hydrogen-rich massive stars generally increase their luminosity and expand during their evolution. As a consequence,
stars above a threshold mass reach the Eddington limit the earlier in their evolution the higher their mass (cf.,
fig.\,5 of Sanyal et al. 2017). For the metallicity of the LMC, inflation occurs in stellar models above $\sim 40\mso$
during late stages of hydrogen burning, and it occurs already at the zero age main sequence for masses above
$\sim 100\mso$. The implication for binary evolution above $\sim 40\mso$ is that all models evolve into
Case\,A mass transfer, i.e., Case\,B does not occur any more. 
Furthermore, the mass donors above $\sim 40\mso$ have an inflated envelope at the onset of Roche-lobe overflow
beyond a limiting initial orbital period which is smaller for higher donor mass.
For hydrogen-free stars with a metallicity of the LMC, inflation occurs above a threshold mass of about 
$24\mso$ (Ishii et al., 1999, K\"ohler et al. 2015, Ro 2019).

The inflated envelope of massive star models is fully convective (Sanyal et al. 2015). 
Furthermore, any mass loss increases the luminosity-to-mass ratio, thus increasing the Eddington
factor. It is therefore not surprising
that Quast et al. (2019) found the mass-radius exponent in such models to be negative (unless steep H/He-gradients
are present in the outermost envelope). They showed that correspondingly, mass transfer due to Roche-lobe overflow
is unstable, like in the case of red supergiant donors. In the absence of more detailed predictions, we therefore
assume that mass transfer with an inflated mass donor leads to a common envelop evolution, and successively
to the merging of both stars, in most cases. %Also in those Case\,A binaries where the mass donor is not inflated at the
%onset of mass transfer, it may develop inflation due to mass stripping if the He-rich core exceeds 24$\mso$.

In the mass-period diagram (Fig.\,\ref{fig:WN}), we have drawn the line beyond which a hydrogen-rich
donor star (assuming here a hydrogen mass fraction of $X=0.4$) would exceed its Eddington limit.
To construct this line, we have used the positions of single star models in the HR diagram 
in which inflation has increased the stellar radius by a factor of two, which coincides 
roughly (fig.\,22 of Sanyal et al., 2015) with the hot edge of the
LBV instability strip (Smith et al. 2004). For a given luminosity on this line, we obtained a corresponding stellar
mass from the mass-luminosity relation of Gr\"afener et al. (2011) for a hydrogen mass fraction of $X=0.4$,
and used the corresponding radius to obtain a binary orbital period for which stars on this line would fill their
Roche-lobe radius for a mass ratio of 0.7. Considering that the orbital period change during Case\,A mass transfer is small 
(Qin et al. 2019), we would not expect to find WR+OB post-mass transfer binaries with H-rich WR stars above this line
if binaries with significantly inflated donor stars would merge. For hydrogen-free Wolf-Rayet stars, the Eddingtion limit 
translates into a simple mass limit, which is also included in Fig.\,\ref{fig:WN}.

In Fig.\,\ref{fig:WN}, we plot the masses and orbital periods of the WN-type binaries in the LMC (Shenar et al. 2019).
We note a group of five massive H-rich short-period WN+O binaries, for which it is unclear whether they did undergo mass transfer 
(cf., Shenar et al. 2019). In any case, they are indeed found below the Eddington limit,
and are thus not in contradiction to having had mass transfer.
The two very massive long-period binaries in Fig.\,\ref{fig:WN}, on the other hand, are clearly pre-interaction systems. 
Even though for lower hydrogen abundances, the line for the H-rich Eddington limit is expected to come down to lower masses, 
the two systems with WN masses just above 30$\mso$ ($\log M_{\rm WN} \simgr 1.5$) show a hydrogen mass fraction of $\sim 0.2$ in the WN star,
for which they would still not violate the Eddington limit. Furthermore, we see that all hydrogen-free WN stars
are located below the corresponding horizontal line. We conclude that the properties of the LMC WN binaries
are in agreement with the assumption that inflated donors lead to mergers.

Since H-free Wolf-Rayet stars may be very close to collapsing to a BH, we add the massive black hole binaries to
Fig.\,\ref{fig:WN} for which the black hole mass is well constrained. Note that we do not include the low- and intermediate
mass black hole binaries here (cf., Casares \& Jonker 2014); their progenitor evolution is not well understood
(Wang et al., 2016). Figure\,\ref{fig:WN} shows that the massive black hole binaries occupy a similar parameter space as
the hydrogen-free WN stars. Figure\,\ref{fig:WN}  can not resolve  whether binaries with initial primary masses above
40$\mso$ contribute to the massive BH-binary population. However, the properties of M33-X7 argue for such a
contribution, since in this binary the BH companion is an O\,star of $\sim 70\mso$. This is not in conflict with
the Eddington limit due to the short orbital period, which implies a progenitor evolution
via Case\,A mass transfer (Valsecchi et al. 2010, Qin et al. 2019).

Nevertheless, Fig.\,\ref{fig:WN} suggests that the contribution of stars above 40$\mso$ to the population
of massive BH-binaries is mostly constrained to orbital periods below $\sim 10\,$d. Therefore, we can
consider the predictions for the number of OB+BH binaries from our Case\,A binary evolution models
as a lower limit, and the corresponding OB star mass distribution for Case A  (Fig.\,\ref{fig:m}) to stretch
out to higher OB masses. Our predictions for larger period OB+BH binaries, which are mostly due to
Case\,B evolution, might not be affected much by this uncertainty.

\subsection{Mass transfer efficiency}

Observations of massive post-mass transfer binaries suggest that the mass transfer efficiency,
i.e., the ratio of the amount of mass accreted by the mass gainer to the amount of mass lost by
the mass donor due to Roche-lobe overflow, is not the same in different binaries. Whereas some can be
better understood with a high mass transfer efficiency,
others require highly non-conservative mass transfer (e.g., Wellstein \& Langer 1999, Langer et al. 2003).
Petrovic et al. (2005) argue for lower efficiency in systems with more extreme mass ratios, and
de Mink et al. (2007) derive evidence for a lower efficiency in wider binary systems.

Our mass transfer model (cf., Sect.\,\ref{sec:method}), which assumes that the mass transfer efficiency drops
when the mass gainer is spinning rapidly, does in principle account for these variations. However, it
requires that sufficient mass is removed from the binary to prevent the mass gainer from exceeding
critical rotation. We apply the condition that the photon energy emitted by the stars in a binary is larger than
the gravitational energy needed to remove the excess material. Otherwise, we stop the model and assume
the binary to merge. Figure\,\ref{fig:grid25} shows the dividing line between the surviving and merging
for our models with an initial primary mass of 25.12$\mso$. %{\em TT: s this related to the boundary line going 45 degrees upward to the right? 
%(between $log P_i=1.0$ and 3.0). In any case, it might be good to explain the slope of this line earlier.} 
The predicted number of OB+BH binaries is
roughly proportional to the area of surviving binaries in this figure.

This condition for distinguishing stable mass transfer from mergers is rudimentary and will need to be replaced
by a physical model eventually. Correspondingly uncertain is the number of predicted OB+BH binaries. However,
Wang et al. (2020) have shown that the distribution of the sizable Be population of NGC\,330 (Milone et al. 2018)
in the colour-magnitude diagram is well reproduced by detailed binary evolution models.
In order to explain their number, however, the condition for stable mass transfer would have to be relaxed
such that merging is prevented in more systems. A corresponding measure would increase the predicted number of OB+BH binaries,
such that, again, our current numbers could be considered as a lower limit.

\subsection{Black hole formation}

As discussed in Sect.\,\ref{sec:method}, our BH formation model is very simple. By applying the single star helium core mass limit
according to simple criteria based on one-dimensional pre-collapse models, and by neglecting small mass ranges
above that limit which may lead to neutron stars rather than black holes, we may overpredict the number of OB+BH systems.
However, the anticipated BH mass distribution is rather flat (Fig.\,\ref{fig:mbh}), such that this overprediction is likely
rather small. Our assumption that the black hole mass equals the final helium core mass is perhaps not very critical,
since it does not affect the predicted number of OB+BH systems. 

The neglect of a BH birth kick may again lead to an overprediction of
OB+BH binaries. However, due to the much larger mass of the BHs compared to neutron stars, birth kicks with similar momenta
as those given to neutron stars upon their formation would still leave most of the OB+BH binaries intact.
While Janka (2013) suggests that NS and BH kick velocities can be comparable in BHs which are produced
by asymmetric fallback, Chan et al. (2018) find only modest BH kicks in their simulations.
{By considering the galactic distribution of low-mass BH binaries, Repetto \& Nelemans (2015)
find 2 out of 7 systems were consisted with a relatively high BH formation kick. This result is confirmed
by Repetto et al. (2017), who find, on the other hand, that the galactic scale hight of the low mass
BH binaries is smaller than that of the low mass NS binaries. Mirabel (2017) provides evidence
for the BHs of $\sim 10\mso$ and $\sim 15\mso$ in the high mass black hole binaries 
GRS\,1915+105 and Cygnus\,X-1 formed with essentially no kick. Furthermore, the systems
which may correspond closest to the our predicted OB+BH distribution, the galactic Be+BH binary
MCW\,656 (Casares et al. 2014) and the potential B+BH binary LB1 (Liu et al. 2019; see our discussion
on this in Sect.\,6), appear to have low eccentricities.}
We consider the systematics of BH kicks to be still open and return to a discussion of their effect on OB+BH systems
in Sect.\,\ref{sec:comtheo}.

\section{Comparison with earlier work}
\label{sec:comtheo}
The computation of large and dense grids of binary evolution models has so far been performed
mostly by using so-called rapid binary evolution codes
(e..g., Hurley et al. 2002, Voss \& Tauris 2003, Izzard et al 2004,
Vanbeveren et al. 2012 , de Mink et al. 2013, Lipunov \& Pruzhinskaya 2014, Stevenson et al 2017, Kruckow et al. 2018).
On the one hand, such calculations can comprehensively cover the initial binary parameter space, and they allow an efficient
exploration of uncertain physics ingredients. On the other hand, stars are spatially resolved by only two grid points, 
{binary interaction products are often described by interpolating in single star models}. 
Therefore, many genuine binary evolution effects are difficult to include, 
which is true for the uncertainties discussed in Sect.\,\ref{sec:uncertain}.

The computation of dense grids of detailed binary evolution models has become feasible in the last decade or so
(Nelson \& Eggleton 2001, de Mink et al. 2007, Eldridge et al. 2008, Eldridge \& Stanway 2016,  Marchant et al. 2016, 2017; see also 
van Bever \& Vanbeveren 1997). Whereas the computational effort is much
larger, detailed calculations are preferable over rapid binary evolution calculations whenever feasible.
Detailed binary model grids have been used to explore various stages and effects of binary evolution,
including the production of runaway stars (Eldridge et al. 2011), double black hole mergers (Eldridge \& Stanway 2016, Marchant et al. 2016)
long-duration gamma-ray bursts (Chrimes et al. 2020), ultraluminous X-ray sources (Marchant et al. 2017),
and galaxy spectra (Stanway \& Eldridge 2019). However, a detailed prediction of the OB+BH binary population has not yet been performed.

Many rapid binary evolution calculations exist. Here, papers predicting OB+BH populations often aim at reproducing the observed
X-ray binary populations (e.g., Dalton \& Sarazin 1995, Tauris \& van den Heuvel 2006,
Van Bever \& Vanbeveren 2000, Andrews et al. 2018). For example, based on the apparent lack of B+BH binaries
in the population of the Galactic X-ray binaries, Belczynski \& Ziolkowski (2009) predicted a very small number of
such systems, using rapid binary evolution models. 
Since the discovery of the massive black hole mergers through gravitational waves, many predictions for the expected
number of double compact mergers have been computed based on rapid binary evolution models (e.g., Chruslinska et al. 2018,
Kruckow et al. 2018, Vigna-Gomez et al. 2018, Spera et al. 2019). However, whereas the binary evolution considered
in these papers includes the OB+compact object stage, their predictions are focused on the double compact mergers.

In the last few years, based on an analytic considerations, Mashian \& Loeb (2017), 
Breivik et al. (2017), Yamaguchi et al. (2018), Yalinewich et al. (2018) and Masuda \& Hotokezaka (2019)
developed predictions for the BH-binary population in the Galaxy. Much of this work concentrates on low-mass MS+BH 
binaries, in view of the currently known 17 low-mass black hole X-ray binaries (McClintock \& Remillard 2006, 
Arur \& Maccarone 2018). Shao \& Li (2019) have recently simulated the Galactic BH-binary population through rapid
binary evolution models, with detailed predictions for OB+BH binaries. As they are largely consistent with the 
outcome of the quoted earlier papers, we compare our results with theirs. 

As shown in Sect.\ref{sec:comobs}, our results imply that the LMC should currently
contain about 120 OB+BH binaries. Assuming a ten times higher star-formation rate in the Milky Way 
(Diehl et al. 2006, Robitaille \& Whitney 2010) would lead to
1200 Galactic OB+BH binaries. Here we neglect the metallicity difference between both systems, which, indeed, for stars
below $40\mso$, is not expected to cause big differences (e.g., Brott et al. 2011), at the level of the accuracy of our consideration.
Shao \& Li exploit the advantage of rapid binary calculations by producing four population models for Galactic MS+BH
binaries, which differ in the assumptions made for the BH kick distribution (see also Renzo et al. 2019). 
They find that essentially no low-mass BH-binaries are produced if efficient BH kicks are assumed.
Given the observed number of low-mass black hole X-ray binaries, Shao \& Li discard the possibility of
efficient BH kicks. For the other cases, they predict between 4\,000 and 12\,000 Galactic OB+BH binaries.
This number exceeds our estimate for the number of Galactic OB+BH binaries by a factor of 3 to 10.

We find three potential reasons for this. First, Shao \& Li adopt a very small accretion efficiency. As in our 
detailed models, they assume that the spin-up of the mass gainer limits the mass accretion. However, in our models,
we check whether the energy in the radiation field of both stars is sufficient to remove the excess material from the
binary system is sufficient, and assume the binary merges if not. No such check is applied by Shao \& Li, with the consequence that
binaries with initial mass ratios as small as 0.17 undergo stable mass transfer. Comparing this with our Fig.\,\ref{fig:grid25}
shows that this might easily lead to a factor of two more OB+BH binaries. Furthermore, Shao \& Li assume that BH can form
from stripped progenitors with masses above 5$\mso$ (we adopted a limit of 6.8$\mso$; see Sect.\,\ref{sec:method}),
and do not discard progenitors with initial primary masses above 40$\mso$ as envelope inflation (see Sect.\,\ref{sec:uncertain})
is not considered in their models. While both effects lead to more OB+BH binaries, they may not be as important as the
first one. 

The distribution of the properties of the OB+BH binaries found by Shao \& Li is similar to those predicted by
our models. The OB stars show a peak in their mass distribution near 10$\mso$, and the BH masses fall in 
the range 5$\dots$15$\mso$ with a peak near 8$\mso$. The orbital periods span from 1 to 1000 days, 
with a peak near $\sim$100\,days, and is similar to that found by Shao \& Li (2014) for Be+BH binaries. 
Naturally, the peak produced by our Case\,A systems (Fig.\,\ref{fig:p}) is not reproduced by the rapid binary evolution models.

%In this respect, we consider it beneficial to look 
%at the evolution of massive binaries half way before the end.
%In OB+BH systems, one star has died, but the other one is still rather unevolved. By stopping here, we avoid the
%accumulation of errors due to additional uncertainties coming into the successive phases of binary evolution on their
%way to the double compact stage, most notably the common envelop evolution, which is particularly uncertain
%for massive binaries (Kruckow et al. 2016).

\section{Comparison with observations}
\label{sec:comobs}

The global H$\alpha$-derived star formation rate of the LMC is about $\sim 0.2\msoy$ (Harris \& Zaritsky, 2009).
About a quarter of that is associated with the Tarantula region, for which the number of
O\,stars is approximately 570 (Doran et al., 2013, Crother 2019). We therefore expect about 2000 O\,stars to be present in
the LMC. About 370 of them have been in the spectroscopic VLT Flames Tarantula survey
(Evans et al. 2011). {Adopting a 3\% probability for a BH companion as suggested by our results
(cf., Sect.\,7)}, we expect about 60 O+BH binaries
currently in the LMC. About 10 of them may have been picked up by the Tarantula Massive Binary Monitoring
survey (Almeida et al. 2017). 

At the same time, we also predict about 1.5\% of the B\,stars above $\sim 10\mso$
to have a BH companion, most of which would likely be Be stars. As they live roughly twice as long as O stars,
and accounting for a Salpeter mass function, we expect about 60 B+BH binaries amongst the $\sim 4000$ B stars
above $10\mso$ expected in the LMC. This means that our models predict more than 100 OB+BH systems in the LMC,
while we know only LMC-X1. The implication is either that our model predictions are off by some two orders of magnitude,
or that the majority of OB+BH binaries are X-ray quiet.

%______________________

   \begin{figure}
   \centering
   \includegraphics[width = 0.8\linewidth, angle=270]{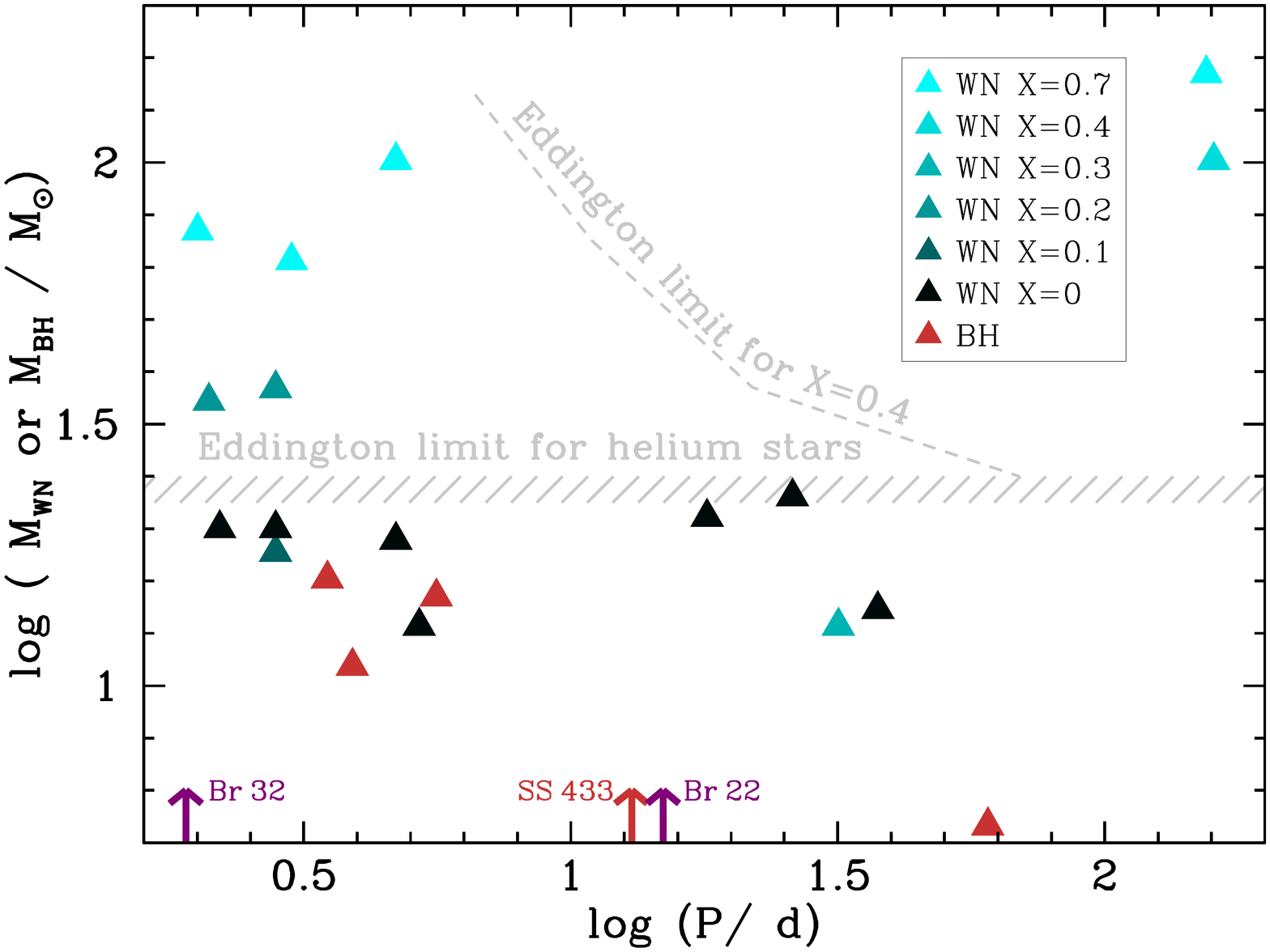}
   \caption{Masses and orbital periods of LMC WN binaries with an O or early B star
   companion (Shenar et al. 2019). The orbital periods of the two LMC WC binaries Br\,22 (right)
   and Br\,32 (left; Boisvert et al. 2008) and of SS\,433 (Hillwig \& Gies 2008) are indicated by arrows.
   Also plotted are the masses and orbital periods of
   the well characterised black holes with O or early B companion, which are, in order of
   increasing orbital period, M33-X7 (Orosz et al. 2007), LMC-X1 (Orosz et al. 2011),
   Cyg-X1 (Orosz et al. 2011), and MCW\,656 (Casares et al. 2014). Above $\sim 24\mso$
   (or a corresponding luminosity of $\llso =5.8$; Gr\"afener et al. 2011), no H-free Wolf-Rayet
   stars are known in the LMC, potentially because this corresponds to their Eddington limit (see text).}
   \label{fig:WN}
   \end{figure}
%______________________

One way to decide which of these two answers is right is to consider the Wolf-Rayet binaries in the LMC.
Shenar et al. (2019) have provided the properties of 31 known or suspected  WN-type LMC binaries. Of those, an orbital period
is known for 16, which we show in Fig.\,\ref{fig:WN}. Of these 16 WN binaries, seven are hydrogen-rich (with hydrogen mass fractions
in the range $0.7 \dots 0.2$), very massive, and likely still undergoing core hydrogen burning.
The other nine are very hot, and most of them hydrogen-free, such that they are likely undergoing core helium burning.
As this implies a short remaining lifetime, they are likely
close to core collapse. In fact, taking their measured mass-loss rates and adopting an average remaining Wolf-Rayet lifetime
of 250\,000\,yr would leave most of them at the end of their lives well above $10\mso$. We can thus assume here that these
nine OB+WN binaries will form OB+BH systems. Once the Wolf-Rayet stars forms a BH, the OB stars will on average still live for a long
time. A remaining OB star lifetime of 1 or 2\,Myr leads to the expectation of 18$\dots$36 OB+BH binaries
currently in the LMC, which is rather close to our model prediction. About 16\% of the 154 Wolf-Rayet stars in the LMC
are of type WC or WO (Breysacher et al. 1999, Neugent et al. 2018). Their properties are less well known; however, at least
three of the 24 WC stars are binaries (the two with well determined orbital period are included in Fig.\,\ref{fig:WN}).
Including the WC binaries will increase the expected number of OB+BH binaries (Sander et al. 2019).

When we look at the properties of the observed WR+OB binaries, we find that the OB\,star masses in the mentioned nine
binaries ($13\dots 44\mso$) are well within the range predicted by our models (Fig.\,\ref{fig:m}). However, 
the average observed OB mass of the nine WR+OB binaries is $\sim 26\mso$, while the average
OB mass of our OB+BH models is about $15\mso$ (Fig.\,\ref{fig:m}). In fact, 
amongst the nine considered LMC systems, only one has a B\,dwarf component (Bat\,29). Of the other potential
WR-binaries listed by Shenar et al. (2019), one more has a B\,dwarf companion but no measured orbital period,
and three more apparently have rather faint B\,supergiant companions  (which is difficult to
understand in evolutionary terms). 
We note that our models predict that the B\,stars 
in such binaries could be rapidly rotating, and that it is unclear whether a Be\,disk 
can be present next to a WR star with a powerful wind. Potentially, the spectral
appearance of B\,stars in this situation may be unusual. Furthermore, 
O\,dwarfs are perhaps easier identified as WR\,star companions than the fainter B\,dwarfs, such that more of the latter could still be discovered.
Another aspect to consider is that a considerable fraction of the
He-star companions of B dwarfs might not have a WR-type spectral
appearance. Their luminosity-to-mass ratio might simply be too low to
yield a sufficient mass-loss for an emission-line spectrum (Sander et
al. 2020, Shenar et al. 2020), eliminating them from being found
in WR surveys

Concerning the orbital periods, a comparison of Fig.\,\ref{fig:p} with Fig.\,\ref{fig:WN} shows that five
of the nine considered WN+OB binaries are found in the period range predicted by our Case\,A binary models,
whereas the other four fall into the Case\,B regime. Notably, the gap in the observed periods ($7\dots 15\,$d)
coincides with the minimum in the predicted period distribution produced between the Case\,A and Case\,B peaks
in the top panel of Fig.\,\ref{fig:p}. On the other hand, our Case\,B models predict a broad distribution
of orbital periods with a peak near 100\,d, whereas the largest measured period is 38\,d (Bat\,64). Again, this
could mean two things. Either our models largely overpredict long-period OB+WN binaries (with core helium burning
WN stars), or many long-period systems have not yet been identified. In this respect, we note that
Shenar et al. (2019) list nine more binaries in which the WR star is likely undergoing core helium burning but
for which no period has been determined. Since longer periods are harder to measure, there could be a bias
against finding long period systems.

This idea is fostered by considering the Be/X-ray binaries. 
This may be meaningful as their evolutionary stage is directly comparable to the OB+BH stage, 
only that the primary star collapsed into a NS, rather than a BH. Because of the larger mass loss 
and the expected larger kick during neutron star formation, in particular the longest period OB+NS 
systems may break-up at this stage, whereas comparable OB+BH systems might survive. 
However, otherwise, we would expect their properties to be quite similar to those of OB+BH systems.
The orbital period distribution of the Galactic Be/X-ray binaries is quite flat and stretches between 10\,d and 500\,d
(Reig 2011, Knigge et al. 2011, Walter et al. 2015).
Overplotted in Fig.\,\ref{fig:p} is the observed orbital period distribution of 24\,Galactic
Be/X-ray binaries following Walter et al.  Figure\,\ref{fig:p} shows that 
the orbital period distribution of the
Be/X-ray binaries matches the prediction of our Case\,B OB+BH binaries very closely.
Since the pre-collapse binary evolution does not know about whether a NS or
BH will be produced by the mass donor, the observed Be/X-ray binary period distribution argues for the existence of
long-period OB+BH binaries as predicted by our models.

Looking at the location of the four massive black hole binaries in the mass-orbital period plot
in comparison to the OB+WR binaries in Fig.\,\ref{fig:WN}, we find that three of them coincide well
with the short-period helium burning WR binaries within the Case\,A range of our models (see also Qin et al., 2019).
Only the Be-BH binary MCW\,656 has a rather large orbital period of 60\,d.
Our conjecture of the existence of many more long-period OB+BH binaries agrees with the anticipation of
Casares et al. (2014), who consider MCW\,656 only as the tip of the iceberg. The reason is that MCW\,656,
in contrast to the short-period OB+BH systems, is X-ray silent, which is likely due to the fact that the wind material
falling onto the BH does not form an accretion disk but an advection-dominated inflow (Shakura \& Sunyaev 1973,
Karpov \& Lipunov 2001, Narayan \& McClintock 2008, Quast \& Langer 2020). We note that also the recently detected
B\,star$-$BH binary system LB-1 (Liu et al. 2019) might fall into this class. 
While it was first proposed that the BH in this system is very massive,
it has been shown subsequently that its mass
is consistent with being quite ordinary (Abdul-Masih et al. 2019b, El-Badry \& Quataert 2020, 
Simon-Dias et al. 2020), if it is a BH at all (Irrgang et al. 2020).

Remarkably, it is the long-period OB+BH binaries which have
the highest chance to produce a double-compact binary which may merge within one
Hubble time.

\section{OB+BH binary detection strategies}
\label{sec:detect}

We have seen above that our binary evolution models predict that about 100 OB+BH binaries remain to
be discovered in the LMC. Scaling this with the respective star-formation rates would lead to 
something like 500 to several thousand OB+BH binaries in the MW. Simplified binary population synthesis models predict similar numbers,
and show that the order of magnitude of the expected number of OB+BH binaries is only weakly dependent on the
major uncertainties in the models (Yamaguchi et al. 2018, Yalinewich et al. 2018, Shao \& Li 2019).  
At the same time, as discussed in Sect.\,\ref{sec:comobs}, the observations of Wolf-Rayet binaries 
and of Be/X-ray binaries, lend strong support to these numbers. Finding these OB+BH binaries, and measuring their properties,
would provide invaluable boundary conditions for the evolution and explosions of massive stars.

One possibility is to monitor the sky position of OB stars and determine the presence of dark companions
from detecting periodic astrometric variations. It has been demonstrated recently that 
the Gaia satellite offers excellent prospects for identifying  OB+BH binaries this way
(Breivik et al. 2017; Mashian \& Loeb 2017; Yalinewich et al. 2018; Yamaguchi et al. 2018, Andrews et al. 2019).
Furthermore, a BH companion induces a photometric variability to an OB star in several ways 
(Zucker, et al. 2007, Masuda \& Hotokezaka 2019).
In the closest  OB+BH binaries, the OB star will be deformed which leads to ellipsoidal variability.
In wide binaries seen edge-on, gravitational lensing of the BH can lead to significant signals (App.\,A). 
Additionally, relativistic beaming due to the orbital motion affects the light curve
of OB+BH binaries. Masuda \& Hotokezaka find that the TESS satellite may help to identify OB+BH binaries,
in particular short-period ones. Finally, OB+BH binaries can be identified spectroscopically,
through the periodic radial velocity shift of the OB component in so-called SB1 systems,
in which only one star contributes to the optical signal. Spectacular examples are provided by
the discovery of the first known Be-BH binary (Casares et al. 2014), 
the potentially similar B[e]-BH binary candidate found by Khokhlov et al. (2018)
and the recently found potential B-BH binary LB-1 (Liu et al 2019; see Sect.\,\ref{sec:comobs}).
Existing surveys include the TMBM survey in the LMC (Almeida et al., 2017), and the Galactic LAMOST 
survey (Yi et al. 2019).
%Giesers B, Dreizler S, Husser T-O, 2018, MNRAS, 475, L15   LMBH

Whichever way the BHs in binary systems affect the signal we are observing from the companion star,
the BH per se will remain unobservable. This means that
the conclusion of having a BH in a given binary will always remain indirect, and --- as physics can never
deliver proofs --- somewhat tentative. This is the more so as, obviously, the technique with which
BH detections are generally associated, namely X-ray observations, appears to fail for the vast majority
of OB+BH binaries (cf., Sect.\,\ref{sec:comobs}). For that reason, it will be beneficial if, firstly,
OB+BH binaries are detected in more than one ways, and secondly, if the properties of the OB component
are measured spectroscopically, to see whether e.g. its surface abundances and its rotation rate fall within expectations.

%\subsection{Comparing OB+BH to OB+OB binaries}
%______________________________________________

   \begin{figure}
   \centering
   \includegraphics[width = \linewidth]{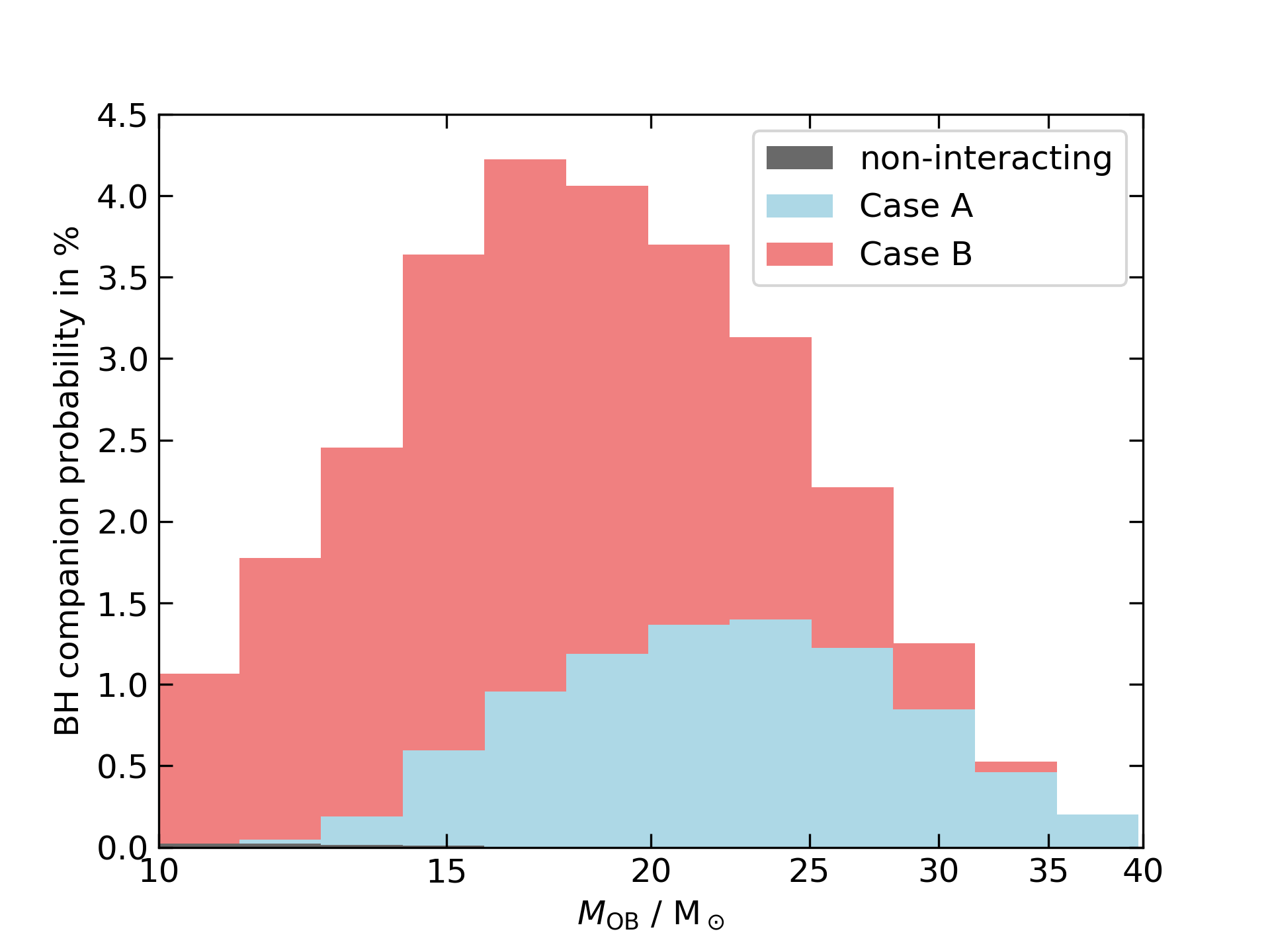}
   \caption{Probability of OB stars of a given mass to have a BH companion, as function of the
   mass of the OB star, according to our population synthesis model. The initial mass function,
   initial binary parameter distributions, and the lifetimes of the OB+BH systems have
   been considered. A initial binary fraction of 100\% has been assumed.}
   \label{fig:mob}
   \end{figure}
%______________________

In our grid of binary evolution models, we produce (potential) OB+BH binaries, but the model systems spend
their largest amount of time as OB+OB binaries. In order to evaluate the probability that a randomly picked OB\,star  
has a BH companion, we divide the number of systems in the mass bin of our OB star
to the corresponding number of OB binaries with any type of companion.
%{\bf Christoph: please one more sentence on how exactly you do this.}
The result is plotted in Fig.\,\ref{fig:mob}. Here, OB single stars are neglected.
Considering them does reduce the probabilities obtained in Fig.\,\ref{fig:mob} by the assumed binary fraction.

Figure\,\ref{fig:mob} resembles the overall OB\,star mass distribution derived in Fig.\,\ref{fig:m}.
However, its ordinate values represent actual BH companion probabilities. Therefore, we find that 
the fraction of OB stars with BH companions is highest in the OB star mass range $14\dots 22\mso$,
with the probability to have an accompanying BH of about 4\%.  For B\,stars near $10\mso$,
the BH companion probability is still about 1\%. For more massive OB stars, we expect BH companions
in at least 1\% of the stars up to about 32$\mso$, where an additional contribution from binaries
with initial primary masses above $40\mso$ is possible (see Sect.\,\ref{sec:uncertain}).

\begin{figure}
   \centering
   \includegraphics[width = \linewidth]{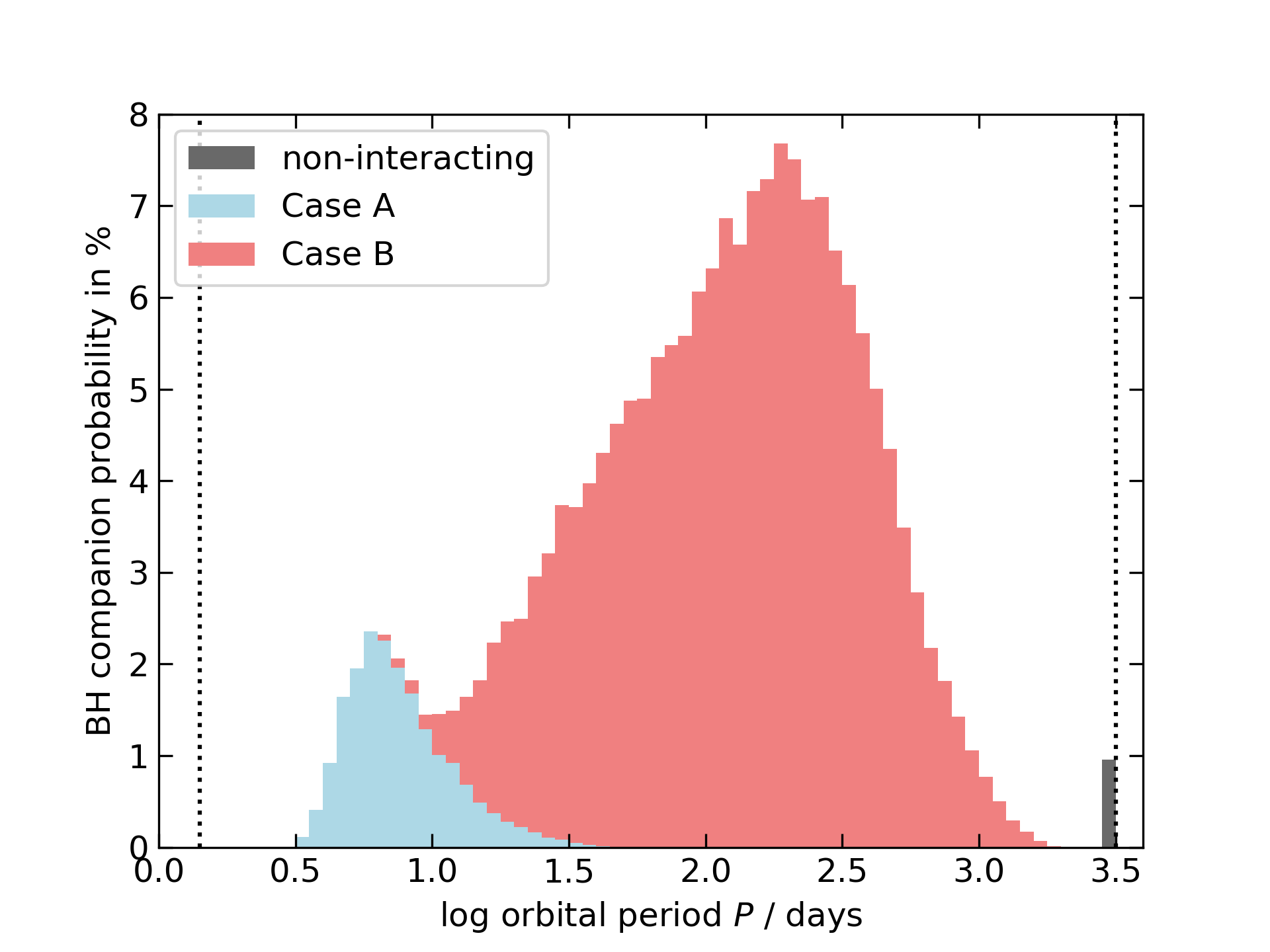}
   \includegraphics[width = \linewidth]{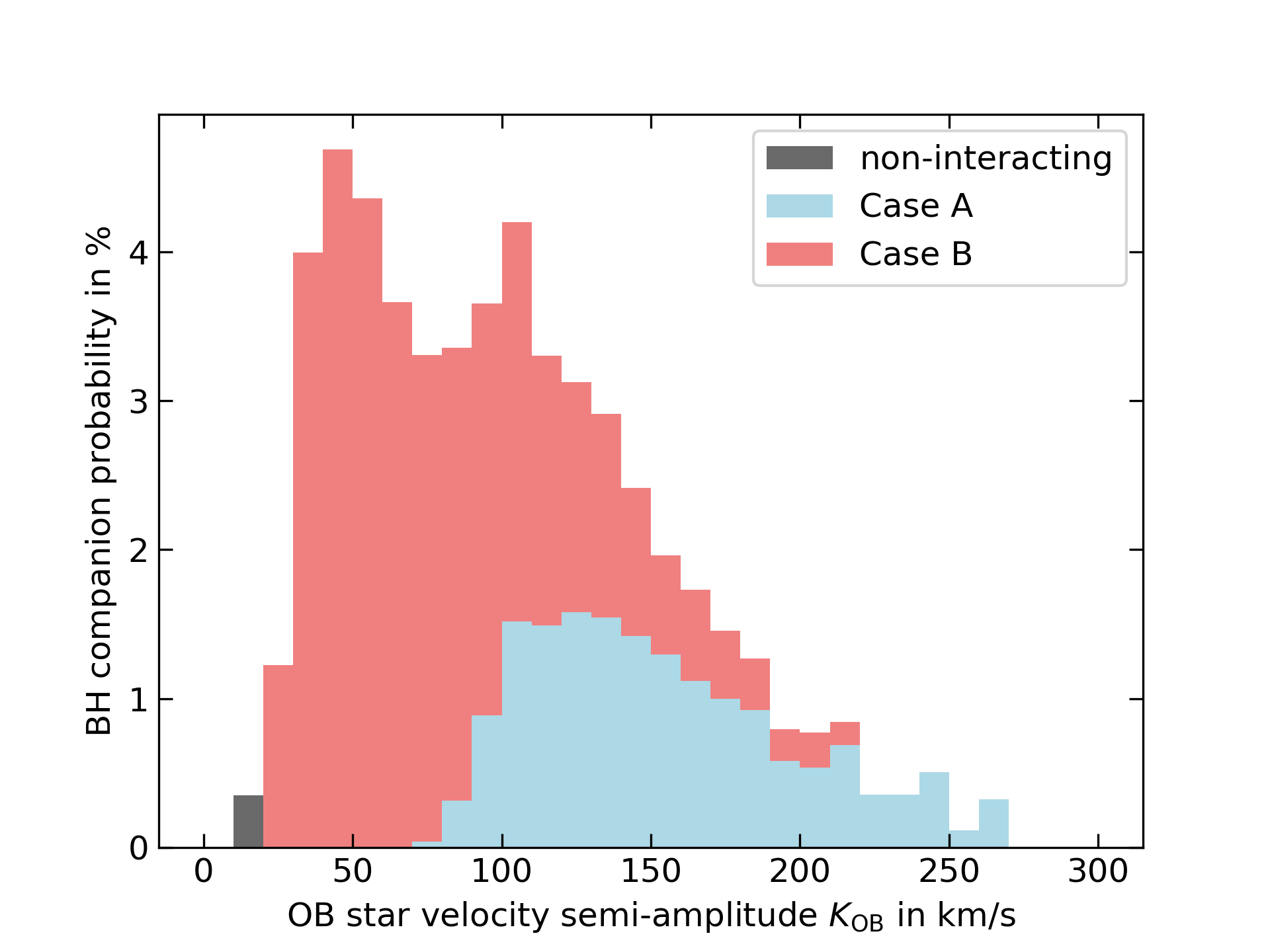}
   \caption{Prediction of our population synthesis model for the probability
   of OB stars to have a BH companion, as function of the observed
   orbital period (top), and the observed radial velocity semi-amplitude
   (bottom), respectively.}
   \label{fig:pk}
   \end{figure}

In the upper panel of Fig.\,\ref{fig:pk}, we show the probability of a randomly picked OB\,binary
to have a BH companion, as function of its orbital period. For example, if our chosen binary has an
orbital period of 10\,d, then its chance to be accompanied by a BH is about 1.5\%. For a period
of 180\,d, on the other hand, it is almost 8\%. Figure\,\ref{fig:pk} shows that the expected orbital 
periods in OB+BH binaries are somewhat ordered according to their initial orbital periods.
The Case\,A systems have the smallest initial periods (cf., Fig.\,\ref{fig:grid25}), and
they produce the shortest period OB+BH binaries in our results. On the opposite side, the initial
period range of the Case\,B binaries is mapped into a quite similar period range of the OB+BH binaries. 

The lower panel of Fig.\,\ref{fig:pk} shows the corresponding distribution of orbital velocities. Again,
the ordinate value in this plot reflects the probability of a randomly picked OB binary to contain a BH,
this time as function of its orbital velocity. The Case\,A systems, which have initial orbital periods
as low as 1.4\,d, provide the fastest moving OB stars, while the Case\,B binaries form many OB+BH systems with
orbital velocities of just a few tens of km/s. 

%______________________

%______________________

%______________________

   \begin{figure}
   \centering
   \includegraphics[width = \linewidth]{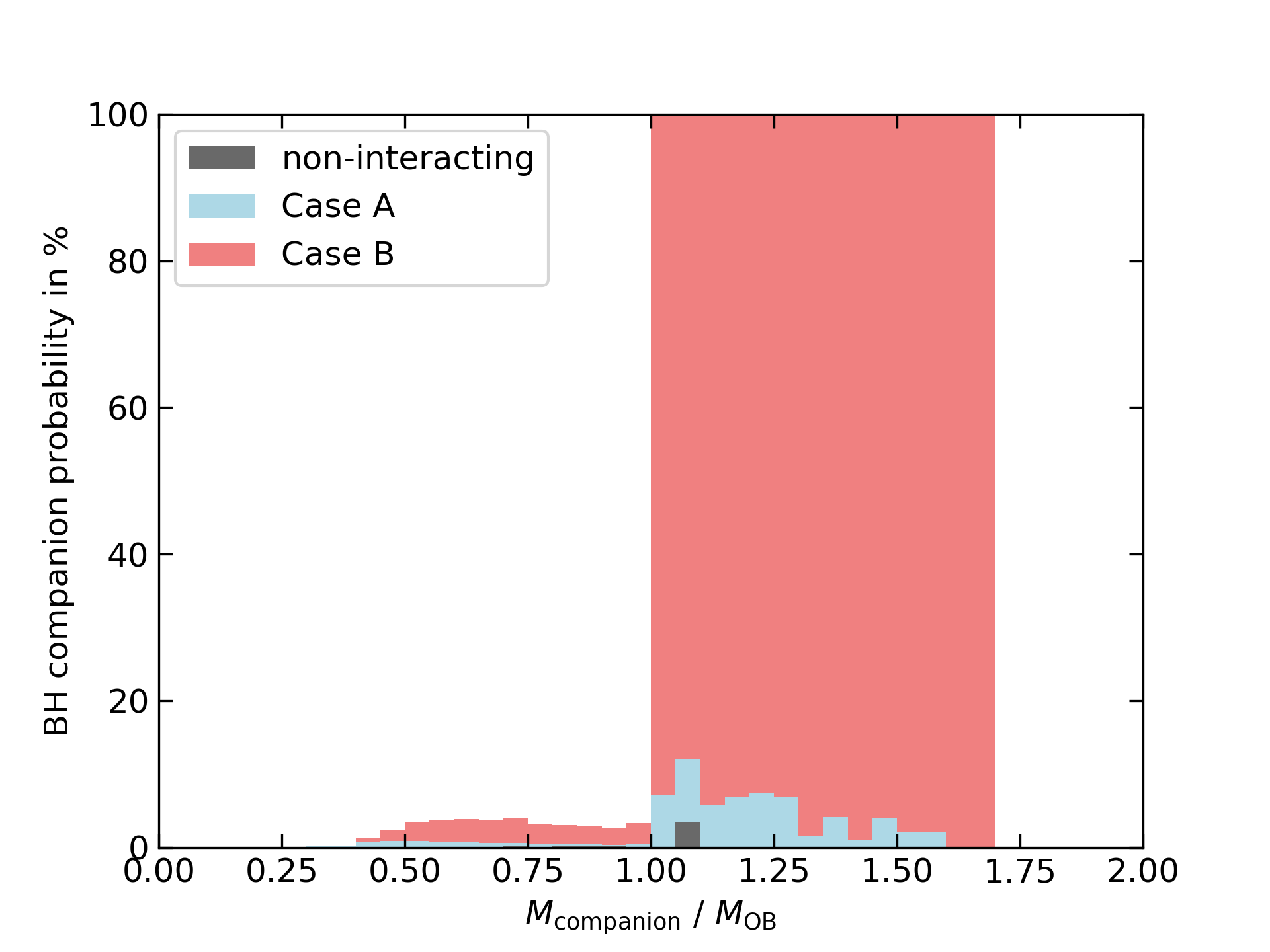}
   \includegraphics[width = \linewidth]{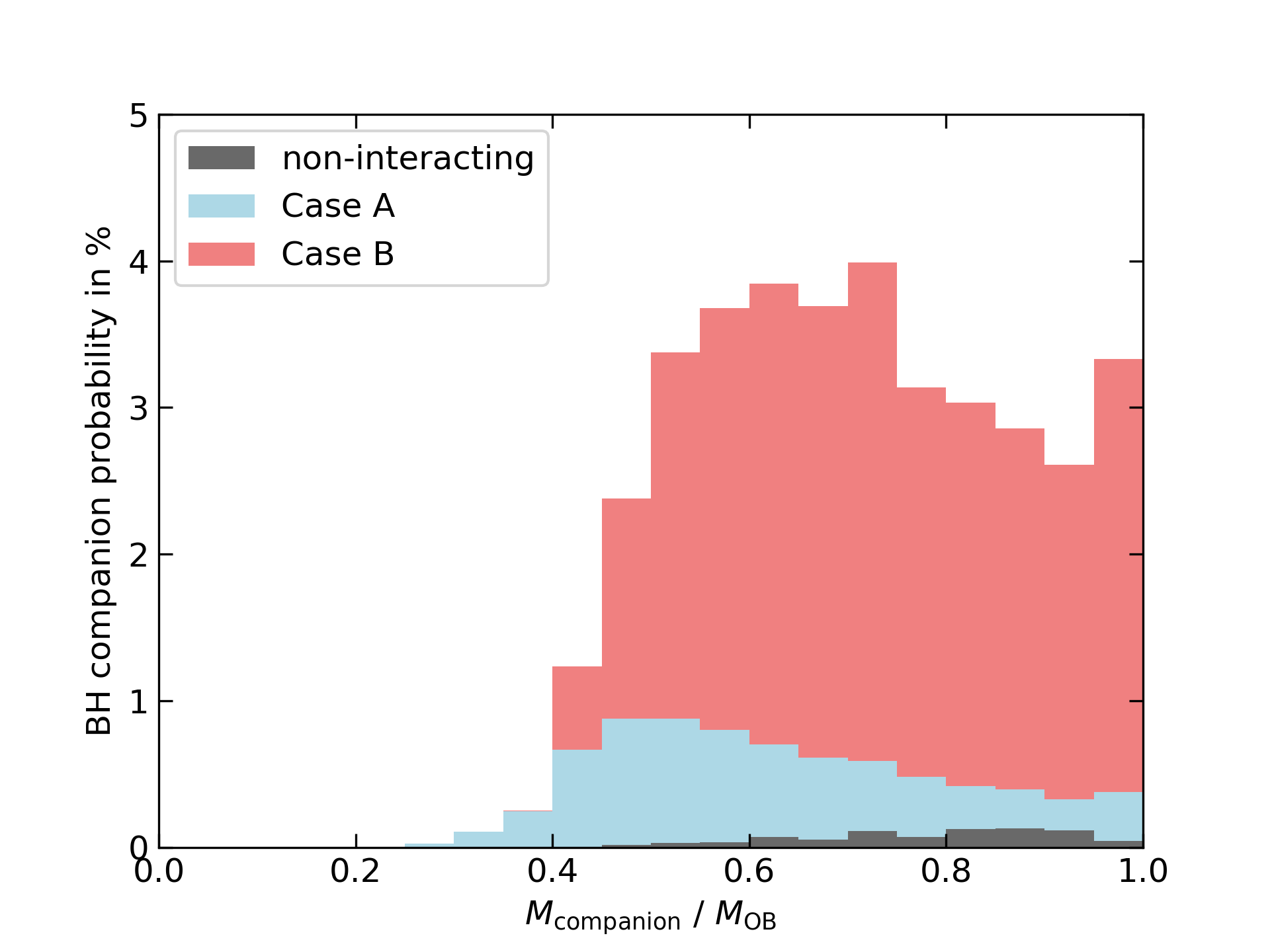}
   \caption{Prediction of our population synthesis model for the probability
   of a randomly picked OB binary to have a BH companion, as function of the mass ratio
   (top). Here, a mass ratio above one means the BH is heavier than the OB star;
   if such an OB binary is picked, its probability of having a BH companion is one.
   The bottom panel shows a zoom of the part with a mass ratio less than one.}
   \label{fig:qq}
   \end{figure}
%______________________

Figure\,\ref{fig:qq} gives the probability of a randomly picked OB binary to be accompanied
by a BH as function of the mass ratio $q=M_{\rm companion}/M_{\rm OB}$. Note that for $q > 1$,
this probability is one. In this case the companion must be a BH and can not be an ordinary star,
since otherwise the ordinary companion star would be the more luminous star of the two, and it would have been
picked as the primary OB star.   

We see that the smallest mass ratios are dominated by Case\,A systems, which is a consequence
of the rather high accretion efficiency in those; i.e., the OB stars in such binaries did gain 
substantial amounts of mass. Combined with Fig.\,\ref{fig:pk}, this means that the OB+BH
binaries with the lowest mass ratios have small orbital periods. 

Finally, Fig.\,\ref{fig:qq} shows that the largest mass ratios produced by our model binaries
is about $q=1.7$. Binaries with such high mass ratios originate from OB+OB binaries
with initially massive primaries and extreme mass ratio, e.g., $40\mso + 13\mso$,
in which the secondary accretes little material. The OB stars in such systems are therefore expected to be
be early B or late O stars. 

%\subsection{Peculiarities of OB stars in OB+BH binaries} 
%\label{sec:ob}

Above, we have discussed the BH companion probabilities of randomly picked OB stars, and found
them to be of the order of a few percent. Considering observing campaigns to search for OB+BH binaries,
an efficiency of a few percent is rather low. However, the OB stars in OB+BH binaries
have had a turbulent life, and signs of that may still be visible. In particular, 
all OB stars in our OB+BH model binaries have accreted some amount of matter from their companion.
As the accretion efficiency in our models drops once the mass gainer has reached critical rotation,
and since a mass increase by about 10\% is required to achieve this (Packet 1981, Petrovic et al. 2005), 
this is roughly the minimum mass increase of our OB mass gainers.

From the properties of the OB stars in OB+BH binaries as described in Sect.\,\ref{sec:results},
we see most OB stars with a BH companion are expected to stand out amongst the
ordinary OB stars. In Case\,A systems, the OB star rotation is expected to be relatively fast, 
but since only the projected rotation velocity can be easily measured, this is not an unambiguous
selection criterion.  However, in our models, the BH companion induces a radial velocity variation 
of 200$\kms$ or more ($K \simgr 100\kms$; Fig.\,\ref{fig:rot}), which should be easily seen even 
though the observed value will again be smaller due to projection (by 21\% on average). In addition,
our models predict a significant surface enrichment with products of hydrogen burning in the fast
majority of all cases, the strongest signature being a clear nitrogen enrichment.

In the Case\,B binaries, surface enrichment of the OB components is predicted to be small. However, their rotation
velocity is expected to exceed 300$\kms$, with values close to critical rotation in those with masses
below $\sim 20\mso$. And even in Case\,B, the expected radial velocity variations of the OB stars 
exceed $40\kms$, with an average well above $100\kms$. 

We note that also the mass ratios of our OB+BH binaries are rather favourable. That means, if
we assume that a MS companion would still be detected as such for mass ratios above 0.5, then
such a companion could be excluded in potential observations of almost all of our OB+BH model binaries.
Given the clues accumulated above, a corresponding search for BHs in SB1 spectroscopic binaries might
thus be promising.

Finally, we want to emphasise that additional possibilities to identify potential OB+BH binaries 
exist when considering the population of young star clusters.
In particular, many of the OB stars in our OB+BH model binaries which evolved via Case\,A mass transfer
have gained a substantial amount of mass. The mass increase may lead the stars 
appear above the cluster turn-off, and the convective core mass increase will rejuvenate them such that
they appear younger than most other cluster stars (van Bever \& Vanbeveren 1997, Schneider et al. 2014, Wang et al. 2020).

\section{Further evolution and connection to double compact mergers} 
\label{sec:future}

%{\em Somewhere in the paper, perhaps in a section (or at least a few paragraphs) elaborate more on the FUTURE destiny of these BH+OB-star binaries. Will they produce TZO-like objects? Will they survive a potential CE? (for the latter, see again Kruckow+2016 on CEs).
%In our joint DNS paper (Tauris+2017), we argued that none (except perhaps one or two) of the one hundred+ known HMXBs would survive the subsequent CE. Can similar statements about the future destiny be made for BH+OB systems? (synthetic and observed).}

As shown in Fig.\,\ref{path}, the OB+BH stage on which we focus here is the last evolutionary stage of massive binaries
which, so far, can be reached with detailed calculations from the ZAMS. Therefore, predictions for later stages 
become more and more uncertain, and are not derived from our models. Nevertheless, it is interesting 
to speculate about the future evolution of the OB+BH. 

First of all, due to the rather large orbital periods of our OB+BH systems (Fig.\,\ref{fig:p}),
in almost all of our model binaries the OB star would fill its Roche-volume only after core hydrogen exhaustion
(Case\,B). We would therefore expect a thermal timescale mass transfer from the OB star to the BH,
with a mass transfer rate of $\dot M \simeq L R / (G M)$.
Since this stage is very short ($\sim 10^4\,$yr), we would expect to observe only very few
systems in this stage, SS\,433 perhaps being one of them (Hillwig \& Gies 2008).
It depends on the mass ejection rate from the mass transferring binary whether a common envelope evolution
is initiated or avoided at this stage. For smaller periods and rather low mass donors, it can perhaps be avoided as estimated
by King et al. (2000) for SS\,433, which has an orbital period of $13.1\,$d and for which a mass ejection rate of
the order of $10^{-4}\,\msoy$ has been determined. For the bulk of our systems, the stellar radius and luminosity will be much larger,
and the mass transfer rate would typically be $10^{-2}\,\msoy$, such that common envelope evolution appears more
likely. With the assumptions for the common envelope evolution as in Kruckow et al. (2016), except for possibly the widest systems, 
we would expect a merging of both stars. 

In any case, the accretion of matter of BHs inside a stellar envelope, and 
the common envelope evolution of a BH and a non-degenerate star, can not yet be
predicted with certainty. Therefore, whether there is a critical orbital period in our predicted
OB+BH period distribution (Fig.\,\ref{fig:p}) beyond which the systems survive the common envelope evolution as a binary,
and what its value would be, remains open. The fact that the peak of the period distribution corresponds to
a rather large value ($\sim 200\,$d) leaves room for the speculation that a significant fraction of the OB+BH binaries
will lead to tight double BH systems.  

\section{Conclusions}
\label{sec:con}

In this paper, we have provided predictions for the properties of the OB+BH binary population in the LMC.
These predictions are based on almost 50\,000 detailed binary evolution models. These models include
internal differential rotation, mass and angular momentum transfer due to Roche-lobe overflow, and no inhibition of
envelope inflation due to the Eddington limit. Only models which undergo stable mass transfer are considered,
implying that common envelope evolution may add additional OB+BH binaries to our synthetic population. 
Our results are subject to substantial uncertainties, which we discuss in detail in Sect.\,\ref{sec:uncertain}.
However, they represent the last long-lived stage of massive binaries on their way to double compact binaries
that can be modeled in detail without interruption starting from the double main sequence stage, which allows the prediction
of their properties with a rather limited amount of assumptions (Sect.\,\ref{sec:method}).
This includes the initially closest binaries which undergo mass transfer during hydrogen burning
(Case\,A), which can be treated only rudimentary in rapid binary evolution calculations. 

We compare our predictions with the number and properties of the observed OB+WR binaries in the LMC, which may be the direct
progenitors of OB+BH binaries. We find good agreement with the mass distribution and with the orbital period distribution
up to $\sim 40\,$d. However, there is a lack of observed long period ($\sim 100\,$d) OB+WR binaries and of 
B+WR binaries, compared to our predictions. While the corresponding observational biases are not well
understood, the similarity of the observed Be/X-ray binary period distribution to that predicted for the OB+BH binaries
argues for the so far undetected presence of long-period unevolved binary companions in a significant fraction of the WR star
population.
  
We derive the distribution of masses, mass ratios, and orbital periods of the expected OB+BH binary population,
and show that OB stars with BH companions may be identified through their radial velocity variations, their
rotation rate or their surface abundances.
Our results imply that an average O or early B\,star in the LMC has a BH companion with a probability of a few percent,
which argues for about 120 OB+BH binaries currently in the LMC. With an about five to ten times higher star-formation rate,
the Milky Way may thus harbour around 1000 of such system. All together, only four such binaries have been found so far,
with one of them in M33.

The vast majority of the predicted OB+BH binaries are expected to be X-ray quiet. The reason is that due to their rather large
expected orbital periods (Fig.\,\ref{fig:pk}), wind material may be accreted in an advection-dominated flow rather than via an
accretion disk. This picture is confirmed by the Be-BH binary MCW\,656 which has an orbital period of 60\,d.
In any case, we have shown that the expected orbital velocities are sufficiently large to identify OB+BH binaries 
spectroscopically (Fig.\,\ref{fig:pk}) --- which is easier here than in their OB+WR progenitors ---,
that the mass ratios are such that main sequence companions can easily be excluded, and that rapid rotation
and/or chemical surface enrichment may help to identify candidate systems.

We find the accumulated evidence for a so far undetected large population of OB+BH binaries significant. 
Its discovery would largely help to reduce the uncertainty in massive binary evolutionary models,
and pave the way to understand the contribution 
of close binary evolution to the BH-merger events observed through their gravitational wave emission.

\begin{acknowledgements}
A.A.C.S. and J.S.V. are supported by STFC funding under grant number
ST/R000565/1. T.M.T.\ acknowledges an AIAS--COFUND Senior Fellowship funded by the European Union’s Horizon~2020 Research and Innovation Programme (grant agreement
no~754513) and Aarhus University Research Foundation. J.B. acknowledges support from the FWO Odysseus program under project G0F8H6N.
A.H. and S.S.D. acknowledge funding by the Spanish Government Ministerio de Ciencia e Innovacion through grants PGC-2018-0913741-B-C22 and SEV 2015-0548, and from the
Canarian Agency for Research, Innovation and Information Society (ACIISI), of the Canary Islands Government, and the European Regional Development Fund (ERDF), under
grant with reference ProID2017010115.
\end{acknowledgements}

\bibliographystyle{aa}
% Cyg X-3: P=4.8h

\begin{appendix}
\label{sec:lens}
\section{Self-lensing of OB+BH binaries}
%lensing (Supermassive BHs: D'Orazio+ 2018MNRAS.474.2975)
The presence of a black hole can potentially be verified by
gravitational lensing magnification. Provided the OB star is
sufficiently well aligned behind the observer--black hole sightline,
the black hole can cause a magnification on the star's flux
(Masuda \& Hotokezaka 2019, D'Orazio \& Stefano 2020).
This lensing magnification would show up as a symmetric peak in the light
curve of the OB star once per orbit.  The maximum magnification is
obtained if star, black hole, and observer are perfectly aligned, and
for star of radius $R_*$ with uniform surface brightness, its value is
$\mu_{\rm max}=\rho^{-1}\sqrt{4+\rho^2}$, where $\rho =R_{\rm OB}/R_{\rm E}$
is the ratio of stellar radius and Einstein radius. Since the distance
of the binary system is much larger than the orbital radius $a$ of the
binary, the Einstein radius for a black hole of mass $M_{\rm BH}$ is
\[
  R_{\rm E}\approx\sqrt{{4GM_{\rm BH}\over c^2}\,a}
  \approx 7.7\times 10^9\,{\rm cm}
  \left( {M_{\rm BH}\over 10\,{\rm M}_\odot} \right)^{1/2}
  \left( {a\over 10^{13}\,{\rm cm}} \right)^{1/2} \;.
\]
Therefore, the dimensionless stellar radius $\rho$ becomes
%
% \left(  \right)
\[
  \rho\approx 65 \left( {R_{\rm OB}\over 5\times 10^{11}\,{\rm cm}}\right)
  \left( {M_{\rm BH}\over 10\,{\rm M}_\odot} \right)^{-1/2}
  \left( {a\over 10^{13}\,{\rm cm}} \right)^{-1/2} \;,
\]
and is thus $\gg 1$. Hence, we can expand the maximum magnification to
yield a maximum brightening of the star by
\begin{eqnarray}
%  |\Delta m|_{\rm max}\!\!\!\!&=&\!\!\!\! 1.086\,\ln\mu_{\rm max}
%  \approx {2.17\over \rho^2}\nonumber \\ \!\!\!\!&\approx&\!\!\!\!
%  5.2\times 10^{-4} \left( {R_{\rm OB}\over 5\times 10^{11}\,{\rm cm}}\right)^{-2}
%  \left( {M_{\rm BH}\over 10\,{\rm M}_\odot} \right)
%  \left( {a\over 10^{13}\,{\rm cm}} \right) \;.
  |\Delta m|_{\rm max}&=& 1.086\,\ln\mu_{\rm max}
  \approx {2.17\over \rho^2}\nonumber \\ &\approx&
  5.2\times 10^{-4} \left( {R_{\rm OB}\over 5\times 10^{11}\,{\rm cm}}\right)^{-2}
  \left( {M_{\rm BH}\over 10\,{\rm M}_\odot} \right)
  \left( {a\over 10^{13}\,{\rm cm}} \right) \;.
  \label{eq:mumax}
\end{eqnarray}

Thus, the maximum brightness increase of the star is of the order of a
milli-magnitude for the fiducial parameters, and scales linearly with
the orbital radius and black hole mass. The magnification decreases
with the misalignment of star, black hole and observer, such that it
drops to about half the value given in Eq.\,(\ref{eq:mumax}) when the
star is misaligned by approximately its own radius. Requiring that the
star passes behind the black hole with a misalignment not larger than
its own radius puts a constraint on the inclination angle $i$ of the
orbital plane of the binary, $\sin(i)\lesssim R_*/a$, or
\[
  i\lesssim 2.85\,{\rm deg}
   \left( {R_{\rm OB}\over 5\times 10^{11}\,{\rm cm}}\right)
   \left( {a\over 10^{13}\,{\rm cm}} \right) ^{-1} \;.
\]
Thus, the orbital plane needs to be well aligned with the sightline to
the binary system in order to yield a brightening larger than $\sim
0.5 |\Delta m|_{\rm max}$.

The prospects for observing lensing magnification in such binary
systems depends sensitively on the photometric accuracy with which the
light curve can be recorded. The lensing nature of the magnification
peaks can be further verified by spectroscopic studies: since the OB
star is predicted to rotate rapidly, the shape of spectral lines
will change during the magnification event, since stellar surface
regions with approaching and receeding (rotational) velocity will be
magnified consecutively. Hence, we expect to see a characteristic time
variability of spectral shapes during the magnification event.

Verifying a lensing event places a strong constraint on the object
causing the lensing -- it has to be smaller than the Einstein radius.

\section{Outcome of the binary models for four additional primary masses}

%______________________________________________

   \begin{figure*}[!htbp]
   \centering
   \includegraphics[width = 0.4\linewidth]{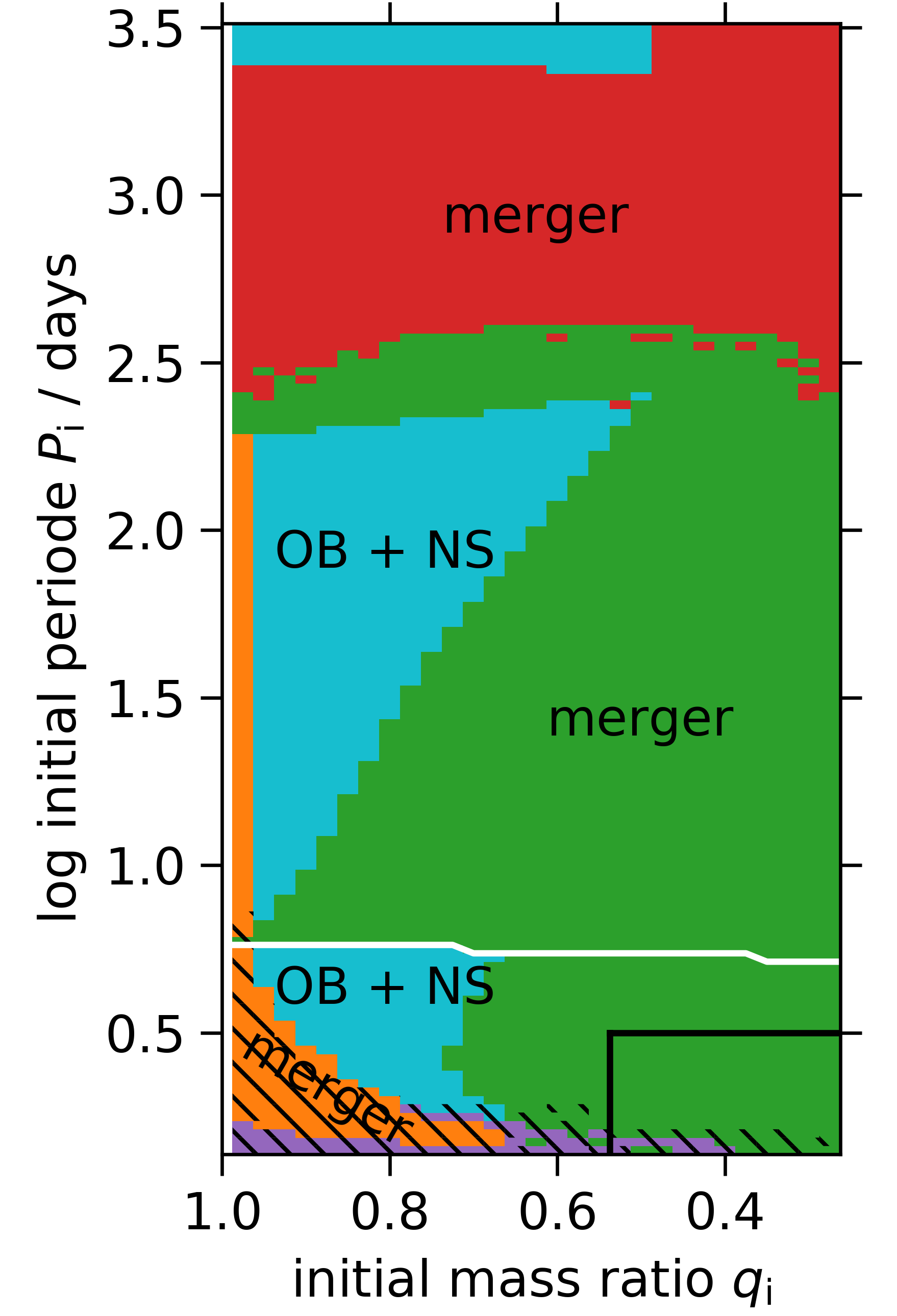}
   \includegraphics[width = 0.4\linewidth]{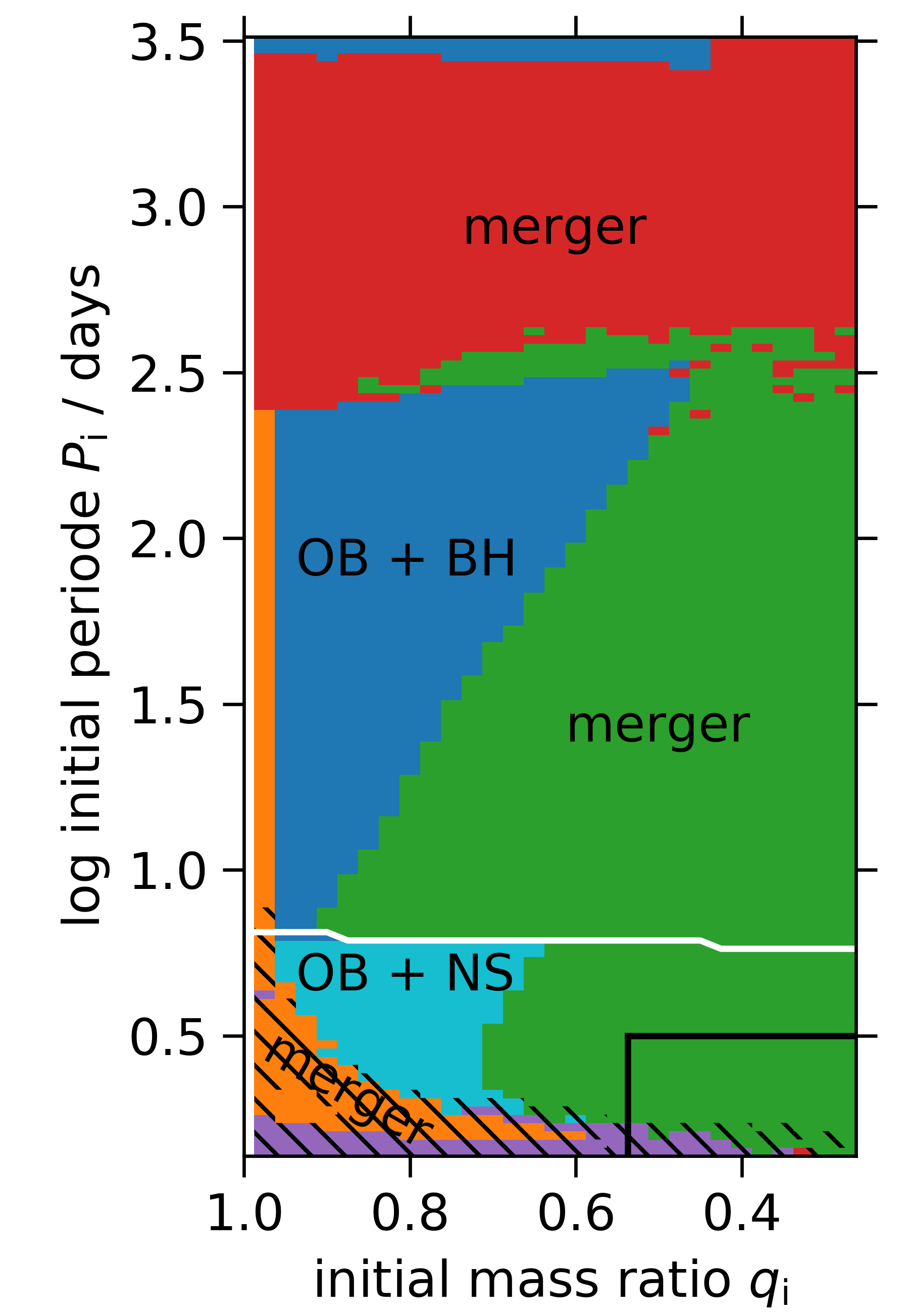}
   \includegraphics[width = 0.4\linewidth]{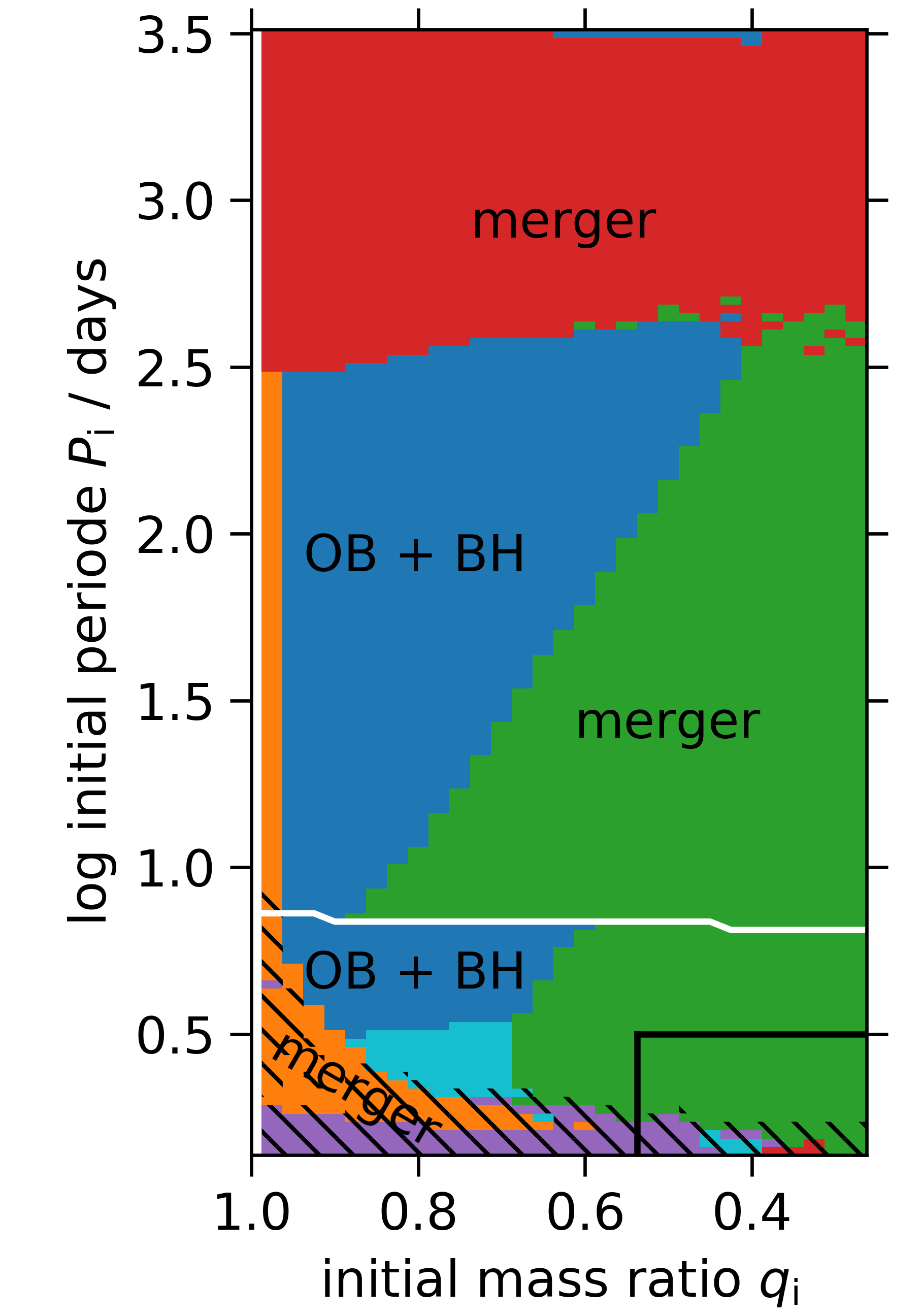}
   \includegraphics[width = 0.4\linewidth]{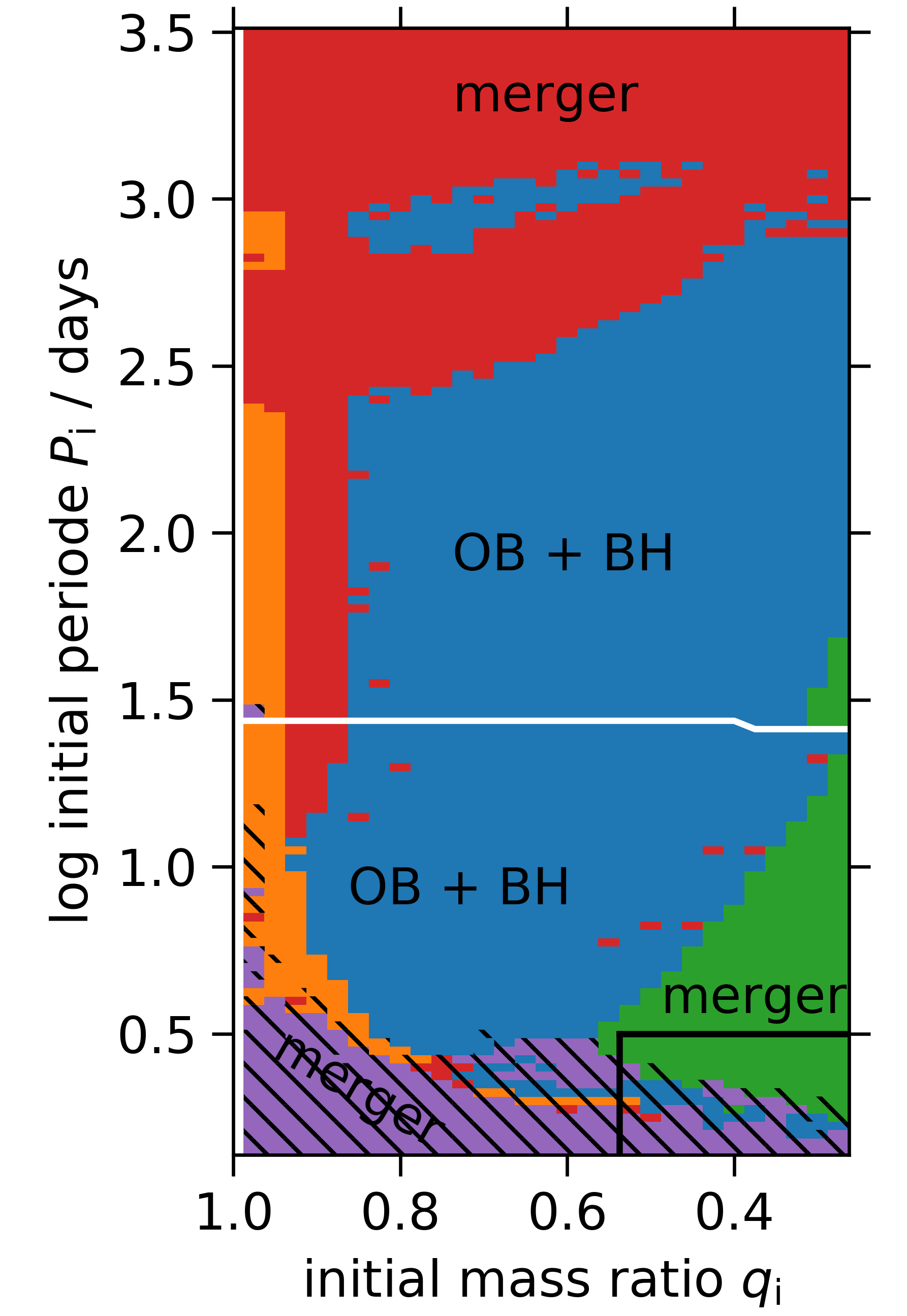}
   \caption{As Fig.\,\ref{fig:grid25}, but for initial primary masses
   of $15.85\mso$ (top left), $17,17\mso$ (top right), 19.19$\mso$ (bottom left) and 39.81$\mso$ (bottom right). 
   The color coding indicates fates as in Fig.\,\ref{fig:grid25}
   (purple: L2-overflow, yellow: inverse mass transfer, green: mass loss limit violation, red:
   common envelope evolution; all assumed to lead to a merger). Black hatching marks contact evolution,
   and the dark blue systems evolve to the OB+BH stage. Here, the light blue color marks systems
   where the mass donor is assumed to form a neutron star rather than a black hole. The white line separates Case\,A and Case\,B
   evolution, and the area framed by the black line in the lower right corner marks the part of the parameter space
   which is disregarded in our results (see Sect.\,\ref{sec:method}). }
   \label{fig:grids}
   \end{figure*}
%______________________

\end{appendix}

\end{document}